\documentclass[prd,twocolumn,aps,letterpaper,amsmath,amssymb,preprintnumbers,showpacs,nolongbibliography,superscriptaddress,floatfix,nofootinbib]{revtex4-2}
\usepackage{array}
\usepackage{graphicx}
\usepackage[export]{adjustbox}
\usepackage[caption=false]{subfig}
\usepackage[normalem]{ulem}
\allowdisplaybreaks 
\usepackage[hypertexnames=true]{hyperref}
\usepackage[capitalize]{cleveref}
\hypersetup{
    colorlinks=true,       
    linkcolor=blue,          
    citecolor=blue,        
    filecolor=blue,      
    urlcolor=blue           
  }

\usepackage{graphicx}
\usepackage{dcolumn}
\usepackage{bm}
\usepackage{xcolor}
\usepackage{slashed}

\usepackage{color}
\definecolor{BrickRed}{rgb}{0.8, 0.25, 0.33}
\definecolor{gray}{rgb}{0.6,0.6,0.6}
\definecolor{darkgreen}{rgb}{0.0, 0.545098, 0.0}
\definecolor{mypink1}{rgb}{0.858, 0.188, 0.478}

\begin{document}
\preprint{FERMILAB-PUB-22-745-T}

\title{Form factor and model dependence in neutrino-nucleus cross section predictions}

\author{Daniel Simons}
\affiliation{Department of Physics and Astronomy, University of Iowa, Iowa City, IA 52242, USA}
\author{Noah Steinberg}
\affiliation{Theoretical Physics Department, Fermi National Accelerator Laboratory, P.O. Box 500, Batavia, IL 60410, USA}
\author{Alessandro Lovato}
\affiliation{Physics Division, Argonne National Laboratory, Argonne, Illinois 60439, USA}
\affiliation{INFN-TIFPA Trento Institute of Fundamental Physics and Applications, Via Sommarive, 14, 38123 Trento, Italy}
\author{Yannick Meurice}
\affiliation{Department of Physics and Astronomy, University of Iowa, Iowa City, IA 52242, USA}
\author{Noemi Rocco}
\affiliation{Theoretical Physics Department, Fermi National Accelerator Laboratory, P.O. Box 500, Batavia, IL 60410, USA}
\author{Michael Wagman}
\affiliation{Theoretical Physics Department, Fermi National Accelerator Laboratory, P.O. Box 500, Batavia, IL 60410, USA}

\begin{abstract}
To achieve its design goals, the next generation of neutrino-oscillation accelerator experiments requires percent-level predictions of neutrino-nucleus cross sections supplemented by robust estimates of the theoretical uncertainties involved. The latter arise from both approximations in solving the nuclear many-body problem and in the determination of the single- and few-nucleon quantities taken as input by many-body methods. To quantify both types of uncertainty, we compute flux-averaged double-differential cross sections using the Green's function Monte Carlo and spectral function methods as well as different parameterizations of the nucleon axial form factors based on either deuterium bubble-chamber data or lattice quantum chromodynamics calculations. The cross-section results are compared with available experimental data from the MiniBooNE and T2K collaborations. We also discuss the uncertainties associated with $N\rightarrow \Delta$ transition form factors that enter the two-body current operator. We quantify the relations between neutrino-nucleus cross section and nucleon form factor uncertainties. These relations enable us to determine the form factor precision targets required to achieve a given cross-section precision.
\end{abstract}

\maketitle

\section{Introduction}

The study of neutrino processes is driven by deep questions whose answers may profoundly change our understanding of physics. In particular, given that neutrinos have mass and mix, the accelerator-neutrino program aims at precisely measuring the parameters that characterize their oscillations, investigating the possible existence of a fourth neutrino flavor, and testing additional Beyond the Standard Model scenarios. 
The success of these experiments rests on our ability to compute neutrino-nucleus cross sections with quantified theoretical uncertainties~\cite{DUNE:2015lol}. The latter presently yield a sizable contribution to the total error budget of oscillation parameters~\cite{T2K:2021xwb,NOvA:2018gge}.

On the other hand, accelerator neutrino experiments allow us access to aspects of nuclear dynamics that would otherwise be difficult to probe at electron-scattering facilities. The chief example is the axial form factor of the nucleon. Its experimental determination dates back to bubble chamber experiments carried out in the 70s and 80s ~\cite{Mann:1973pr,Barish:1977qk,Baker:1981su,Miller:1982qi,Kitagaki:1983px} and to electroweak single pion production measured at ANL and BNL in the 80s ~\cite{Barish:1978pj,Radecky:1981fn,Kitagaki:1986ct,Kitagaki:1990vs}. A simple dipole parameterization with axial mass $M_A\sim 1$ GeV reproduces single-nucleon data. In contrast, a larger value $M_A\sim 1.2$ GeV was required in order to make relativistic Fermi Gas predictions for the neutrino-$^{12}$C cross sections compatible with MiniBooNE data~\cite{MiniBooNE:2010bsu}.
Note that the spectral-function formalism, which includes the vast majority of nuclear correlations, requires an even larger value of $M_A$ to reproduce experimental data~\cite{Benhar:2009wi}. 

This apparent inconsistency appeared to be solved by models that include two-body current operators, as they can reproduce MiniBooNE and T2K data with $M_A~\sim~ 1$~GeV~\cite{Martini:2010ex,Martini:2011wp,Nieves:2011pp,Nieves:2011yp,Megias:2014qva,Gonzalez-Jimenez:2014eqa,Pandey:2014tza}. However, these models are somewhat simplified, as they are based on a mean-field description of nuclear dynamics. As in the one-body current case, it may well be that when nuclear correlations are accounted for there is room for larger values of $M_A$. This is confirmed by recent Green's function Monte Carlo (GFMC) results, which provide a full account of nuclear correlations and two-body current effects~\cite{Lovato:2020kba}. However, the non-relativistic nature of the GFMC hampers its applicability to neutrino accelerator experiments where the neutrino flux energy is of the order of a few GeV. The spectral function (SF) method, relying on the factorization of the final hadronic state, allows for the inclusion of relativistic effects and exclusive channels while retaining most of the important effects coming from multi-nucleon dynamics. The factorization scheme has been extended to include one- and two-body current operators in a consistent fashion as well as pion production amplitudes and validated against electron scattering data~\cite{Rocco:2018mwt,Rocco:2019gfb}. The spectral function of light and medium mass nuclei has been recently computed exploiting quantum Monte Carlo (QMC) techniques and shares with the GFMC the same description of nuclear dynamics ~\cite{Andreoli:2021cxo,CLAS:2022wvz}. 
Comparing the results obtained for lepton-nucleus scattering using these two different approaches enables a precise quantification of the uncertainties inherent to factorization schemes that need to be accounted for when assessing the total error of the theoretical calculations 
in neutrino oscillation analysis. 

There has been significant recent progress in lattice quantum chromodynamics (LQCD) calculations of nucleon axial and vector form factors~\cite{Shintani:2018ozy,Jang:2019vkm,Jang:2019jkn,RQCD:2019jai,Alexandrou:2020okk,Ishikawa:2021eut,Meyer:2021vfq,Park:2021ypf,Djukanovic:2022wru}, which have now been performed using approximately physical values of the quark masses as well as multiple lattice spacings and volumes to enable continuum, infinite-volume extrapolations~\cite{RQCD:2019jai,Park:2021ypf,Djukanovic:2022wru}. 
Vector form factor results show encouraging consistency between LQCD and experimental determinations~\cite{Alexandrou:2018sjm,Jang:2019jkn,Park:2021ypf}.
Excited-state effects involving $N\pi$ states have been identified as a significant source of systematic uncertainty in LQCD calculations of axial form factors~\cite{Bar:2018xyi,Bar:2019gfx,Jang:2019vkm,RQCD:2019jai,Park:2021ypf,Barca:2021iak}. More sophisticated analysis strategies employed in recent calculations~\cite{Jang:2019vkm,RQCD:2019jai,Park:2021ypf}, as well as variational methods used to study $\pi\pi$~\cite{HadronSpectrum:2009krc,Morningstar:2011ka,Dudek:2012xn,Dudek:2013yja,Wilson:2015dqa}, $N\pi$~\cite{Lang:2012db,Kiratidis:2015vpa,Andersen:2017una,Silvi:2021uya,Bulava:2022vpq}, and $NN$~\cite{Francis:2018qch,Horz:2020zvv,Amarasinghe:2021lqa} scattering and $N\pi$ transition form factors~\cite{Barca:2021iak} that enable $N\pi$ and other excited-state effects to be explicitly subtracted from future nucleon elastic form factor calculations, provide paths towards better quantifying and reducing these challenging systematic uncertainties.
Although significant future progress is expected, LQCD calculations of nucleon axial form factors have progressed to a point where it is important to understand the phenomenological impact of current results and quantitatively establish what form factor precision is required to achieve the cross-section precision needs of current and future neutrino oscillation experiments~\cite{Kronfeld:2019nfb,Meyer:2022mix,Ruso:2022qes}.

Current LQCD results predict a significantly larger axial form factor at $Q^2 \sim 1$~GeV$^2$ than determinations from deuterium bubble chamber data.\footnote{A similar trend has recently been obtained within continuum Schwinger function methods~\cite{Chen:2021guo,ChenChen:2022qpy}.} 
LQCD form factor results were incorporated into the GENIE~\cite{Andreopoulos:2009rq} neutrino event generator and shown to lead to significant differences in neutrino-nucleus cross sections in Ref.~\cite{Meyer:2022mix}. The interplay of these form factor changes with other aspects of GENIE remains to be studied, and in particular the ``empirical MEC model'' of two-body currents in GENIE and tuned versions of this model used in neutrino oscillation experiments~\cite{NOvA:2020rbg,GENIE:2021zuu,GENIE:2022qrc} are obtained by fitting cross-section predictions to data while assuming a particular axial form factor model for quasi-elastic contributions. It is important to understand axial form factor effects on the event generators used to analyze current experiments, but using such a data-driven approach to two-body current contributions makes it difficult to disentangle the effects and uncertainties associated with nucleon form factors from those associated with two-body currents and other nuclear effects.

The theoretically well-defined separation between one- and two-body currents in the GFMC and SF results of this work, in conjunction with the study of nuclear model dependence enabled by comparing two realistic many-body methods, enables robust quantification of the effects and uncertainties of nucleon axial form factors and other single-nucleon inputs to neutrino-nucleus cross sections. Parameterizing the nucleon axial form factor using the model-independent $z$ expansion~\cite{Hill:2006ub,Hill:2010yb,Bhattacharya:2011ah,Meyer:2016oeg} allows uncertainty quantification to be performed without introducing any dependence on assumptions about the shape of the axial form factor.
Relations between neutrino-nucleus cross section uncertainties and $z$ expansion parameter uncertainties are determined below and used to construct quantitative precision targets for future LQCD calculations of nucleon axial form factors. The effects of  $N\rightarrow \Delta$ transition form factors uncertainties are also studied.

This paper is organized as follows. Section~\ref{sec:methods} introduces the formalism necessary to compute the inclusive neutrino-nucleus cross section. Section~\ref{sec:axialFF} is dedicated to the determination of the axial form factor and its uncertainties estimation. Results are presented in Section~\ref{sec:results}, while Section~\ref{sec:conclusion} provides concluding remarks and discusses the outlook for future work.

\section{Methods}
\label{sec:methods}

In the one-boson exchange approximation, the neutrino-nucleus double differential cross sections can be written in the form
\begin{equation}
\begin{aligned}
    \left(\frac{d\sigma}{dT_{\mu}d\cos\theta_{\mu}}\right)_{\nu/\bar{\nu}} &= \frac{G^2}{2\pi}\frac{k'}{2E_{\nu}}[L_{CC}R_{CC}+ 2L_{CL}R_{CL} \\
    & + L_{LL}R_{LL} + L_{T}R_{T} \pm 2L_{T'}R_{T'}],
    \label{eq:cross_sec}
\end{aligned}
\end{equation}
 where the $\pm$ sign corresponds to $\nu(\bar{\nu})$ scattering. We take $G = G_{F}\cos\theta_{c}$, with $G_{F} = 1.1664\times10^{-5}\,\rm{GeV^{-2}}$ ~\cite{Nakamura:2002jg} and $\cos\theta_{c} = 0.9740$ ~\cite{ParticleDataGroup:2010dbb} for charged current processes studied here. The $L_{i}$ factors only depend upon the kinematics of the initial and final state leptons, while the electroweak response functions $R_{i}$ are defined as linear combinations of different components of the hadronic response tensor $R^{\mu\nu}$
\begin{equation}\label{eq:ResponseFunctionE}
    R^{\mu\nu} = \sum_{f}\langle 0|J^{\mu\dagger}|f\rangle\langle f|J^{\nu}|0\rangle\delta (E_{0} + \omega - E_{f}).
\end{equation}
where $|0\rangle$ is the nuclear ground state, $|f\rangle$ are all possible final states of the $A$-nucleon system and $J^{\mu}$ is the nuclear current operator.
The response tensor contains all the information on the structure of the nuclear target, defined in terms of a sum over all transitions from the ground state to any final state, including states with additional hadrons. 
In this work we consider the one- and two-body contributions to the nuclear current operator
\begin{equation}
    J^{\mu} = \sum_{i}j^{\mu}_{i} + \sum_{j>i}j^{\mu}_{ij}. 
\end{equation}
and we will exclusively focus on final states involving only nucleons, ignoring single-nucleon excitation processes. Exact expressions for $L_{i}$ and $R_{i}$ can be found in Ref.~\cite{Shen:2012xz}. 
\vskip 0.12in
Computing the electroweak response functions in the energy regime relevant to oscillation experiments is a highly-nontrivial task. Existing approaches  
differ based on approximation schemes and kinematic regimes of applicability. In this work, we will consider two different methods that model the initial target state in a similar fashion, but differ in the treatment of the interaction vertex and final-state interactions: the Green's function Monte Carlo, and the spectral function approaches. 

\subsection{Green's Function Monte Carlo}
\label{sec:GFMC}
The response function in Eq.~\eqref{eq:ResponseFunctionE} is expressed in the energy domain.
It can be related to a matrix element of electroweak currents in the time domain,
\begin{eqnarray}
    \label{resp}
    R_{\alpha\beta}(\mathbf{q},\omega) &  = & \int
    dt e^{i (\omega+E_0) t} \  \langle 0 | J_{\alpha}^\dagger \ e^{- i H t}\  J_\beta | 0 \rangle,
    \label{eq:ResponseFunctionT}
\end{eqnarray}
where a completeness relation has been inserted to carry out the sum over the possible final states of the $A$-nucleon system. The non-relativistic nuclear Hamiltonian $H$ used in these calculations is comprised of the Argonne $v_{18}$~\cite{Wiringa:1994wb} (AV18) nucleon-nucleon potentials plus the Illinois-7~\cite{Pieper:2008rui} (IL7) three-nucleon force. The Green's function Monte Carlo method uses an imaginary-time projection to distill information on the energy dependence of the response functions. In particular, the imaginary-time response is obtained by replacing the real time entering  Eq.~\eqref{eq:ResponseFunctionT} with the imaginary time $t \to -i \tau$.
The Laplace transform of the energy-dependent response, called the Euclidean response function, is defined by
\begin{equation}
  E_{\alpha\beta}(\mathbf{q},\tau) = \int_{\omega_{\rm th}}^\infty d\omega\, e^{-\omega \tau} R_{\alpha\beta} (\mathbf{q},\omega)\,,
\end{equation}
where $\omega_{\rm th}$ is the inelastic threshold. 
Bayesian techniques, most notably maximum entropy~\cite{Bryan:1990,Jarrell:1996rrw}, can be used to retrieve the energy dependence of the response functions from their Laplace transform. Following this strategy, GFMC calculations have been successfully performed to obtain the inclusive electroweak response function of light nuclei while fully retaining the complexity of many-body correlations and associated electroweak currents~\cite{Lovato:2016gkq,Lovato:2017cux,Lovato:2020kba}.
Recently, algorithms based on artificial neural networks have been developed to invert the Laplace transform~\cite{Raghavan:2020bze} and show better accuracy that maximum entropy, especially in the low-energy transfer region.

The five response functions entering the CC cross section have been calculated with GFMC methods for momentum transfers in the range (100--700) MeV in steps of 100 MeV. In order to compute the flux-averaged cross sections presented in Sec.~\ref{sec:results} that involve larger momentum transfers, we capitalize on the scaling features of the GFMC response functions discussed in Refs.~\cite{Rocco:2018tes,Lovato:2020kba,Barrow:2020mfy} to interpolate and safely extrapolate them at momentum transfers $|\mathbf{q}| > 700$ MeV. The charge-changing weak current is the sum of vector and axial components
\begin{align}
    J^{\mu} & = J^{\mu}_{V} - J^{\mu}_{A}, \nonumber \\ J_\pm & =J_x\pm i J_y\, .
\end{align}
These operators comprise both one- and two-body contributions, whose expressions are reported in Ref.~\cite{Shen:2012xz}. Here, we focus on the axial and pseudo-scalar one-body contributions given by 
\begin{align}
  j^0_A &= -\frac{F_A}{2m_N}\tau_\pm \bm{\sigma}\cdot \{{\bf k},e^{i{\bf q}\cdot{\bf r}}\} \nonumber\\
   {\bf j}_A & = -F_A \tau_\pm\Big[ {\bm \sigma}e^{i{\bf q}\cdot{\bf r}}-\frac{1}{4m_N^2}\Big( {\bm \sigma}\{{\bf k}^2,e^{i{\bf q}\cdot{\bf r}}\}
   - \{ ({\bm \sigma}\cdot {\bf k}){\bf k}, e^{i{\bf q}\cdot{\bf r}}\}\nonumber\\
    &  -\frac{1}{2}{\bm \sigma\cdot {\bf q}} \{{\bf k},e^{i{\bf q}\cdot{\bf r}}\} -\frac{1}{2} {\bf q}\{ ({\bm \sigma}\cdot {\bf k}), e^{i{\bf q}\cdot{\bf r}}\}+i {\bf q}\times {\bf k}e^{i{\bf q}\cdot{\bf r}} \Big) \Big] \nonumber\\
    j^\mu_{P} & = \frac{F_{P}}{2m_N^2} \tau_\pm q^\mu {\bm \sigma}\cdot {\bf q}e^{i{\bf q}\cdot{\bf r}}\, ,
\end{align}
where $m_N$ is the nucleon mass and $\{\cdot,\cdot\}$ denotes the anticommutator.
The parametrization of the axial form factor $F_{A}$ has historically involved  models with and without over-constraining theoretical assumptions and will be a focus of this work in Sec.~\ref{sub_sec:fa_param}. 
For the pseudo-scalar form factor, PCAC and pion-pole dominance relations valid in leading order chiral perturbation theory~\cite{Ber94,Bernard:2001rs} can be used to relate $F_{P}$ to $F_{A}$ as
\begin{equation}\label{eq:PCAC}
F_{P} = \frac{2F_{A}m_{N}^2}{m^2_{\pi} + Q^2},
\end{equation}
where $Q^2 = -q^2 = -(q^0)^2 + \mathbf{q}^2$ and $m_\pi$ is the pion mass.
This relation is consistent with the pseudoscalar form factor constraints extracted from precise measurements of the muon-capture rate on hydrogen
and $^3$He~\cite{And07-all}. It is also consistent with current LQCD results~\cite{RQCD:2019jai,Park:2021ypf}. Precise future LQCD calculations will further test Eq.~\eqref{eq:PCAC} and provide independent determinations of $F_P$ without theoretical assumptions on its relation to $F_A$.

The isovector two-body currents can be separated into model-dependent and -independent terms. Those associated with pion exchange are denoted as model independent; they are constrained by the continuity equation and do not contain any free parameters, since they are determined directly from the nucleon-nucleon interaction. The expressions adopted in this work and reported in Refs.~\cite{Shen:2012xz,Carlson:1997qn} are consistent with the semi-phenomenological AV18 interaction. The diagrams associated with the intermediate excitation of a $\Delta$ isobar are purely transverse and generally referred to as model dependent. Their expression can be found in Refs.~\cite{Shen:2012xz,Carlson:1997qn}, it is important to mention that in order to be used in the GFMC calculations, the static $\Delta$ approximation has to be adopted, i.e. the kinetic-energy contributions in the denominator of the $\Delta$ propagator are neglected. The impact of this approximations will be further discussed when comparing the GFMC and SF results. 
Among the axial two-body current operators, the leading terms of pionic range are those associated with excitation of $\Delta$-isobar resonances, despite being less relevant the two-body operators associated with axial $\pi NN$ contact interactions are also accounted for, we 
refer the reader to Refs~\cite{Shen:2012xz,Carlson:1997qn} for a more detailed discussion of these contributions.

\subsection{Extended Factorization Scheme}
\label{sec:FactorizationScheme}
At large values of momentum transfer ($|\textbf{q}| \gtrsim 400$ MeV), a factorization scheme can be used in which the neutrino-nucleus scattering is approximated as an incoherent sum of scatterings with individual nucleons, and the struck nucleon system is decoupled from the rest of the final state spectator system. 
In the quasielastic region, the dominant reaction mechanism is single nucleon knockout. In this case, the final state is factorized according to
\begin{equation}
    |f\rangle = |\textbf{p}'\rangle \otimes |\Psi^{A-1}_{f},\textbf{p}_{A-1}\rangle,
    \label{eq:fact:1b}
\end{equation}
where $|\textbf{p}'\rangle$ is the final state nucleon produced at the vertex, assumed to be in a plane wave state, and $|\Psi^{A-1}_{f},\textbf{p}_{A-1}\rangle$ describes the residual system, carrying momentum $\textbf{p}_{A-1}$.
\vskip 0.12in
Inserting this factorization ansatz as well as a single-nucleon completeness relation gives the matrix element of the 1 body current operator as
\begin{equation}\label{eq:1bodyME}
    \langle f| j^{\mu} |0\rangle \rightarrow \sum_{k}[\langle \Psi^{A-1}_{f}|\otimes\langle k|] |0\rangle\langle p|\sum_{i}j^{\mu}_{i}|k\rangle,
\end{equation}
where $\textbf{p} = \textbf{q} + \textbf{k}$. This first piece of the matrix element explicitly does not depend on the momentum transfer and so can be computed using techniques in nuclear many body theory. The second piece can be straightforwardly computed once the currents $j^{\mu}_{i}$ are specified as the single nucleon states are just free Dirac spinors. Substituting the last equation into Eq.~\eqref{eq:ResponseFunctionE}, and exploiting momentum conservation at the single nucleon vertex, allows us to rewrite the one body contribution to the response tensor as
\begin{equation}\label{eq:W1body}
\begin{aligned}
    R_{1b}^{\mu\nu}(\textbf{q},\omega) = &\int\frac{d^{3}k}{(2\pi)^3}dEP_{h}(\textbf{k},E)\frac{m^{2}_{N}}{e(\textbf{k})e(\textbf{k} + \textbf{q})} \\ 
    & \times\sum_{i}\langle k|j_{i}^{\mu\dagger}|k +q\rangle\langle k+q|j_{i}^{\nu}|k\rangle \\ 
    & \times\delta(\tilde{\omega} + e(\textbf{k}) - e(\textbf{p})),
\end{aligned}
\end{equation}
where $e(\mathbf{k}) = \sqrt{m_N^2 + \mathbf{k}^2}$.
The factors $m_{N}/e(\textbf{k})$ and $m_{N}/e(\textbf{k} + \textbf{q})$ are included to account for the covariant normalization of the four spinors in the matrix elements of the relativistic current. The energy transfer has been replaced by $\tilde{\omega} = \omega - m_{N} + E - e(\textbf{k})$ to account for some of the initial energy transfer going into the residual nuclear system. Finally, the calculation of the one-nucleon spectral function $P_{h}(\textbf{k},E)$ provides the probability of removing a nucleon with momentum $\textbf{k}$ and leaving the residual nucleus with an excitation energy $E$. Its derivation using QMC techniques is discussed in Sec.~\ref{sub_sec:sf}. 
\vskip 0.12in
The relativistic current used in Eq.~\eqref{eq:W1body} is given as a sum of vector and axial terms which can be written as
\begin{equation}\label{eq:FF}
\begin{aligned}
    &j^{\mu}_{V} = F_{1}\gamma^{\mu} + i\sigma^{\mu\nu}q_{\nu}\frac{F_{2}}{2m_{N}} \\
    &j^{\mu}_{A} = \gamma^{\mu}\gamma^{5}F_{A} + q^{\mu}\gamma^{5}\frac{F_{P}}{m_{N}}.
\end{aligned}
\end{equation}
In the above, $F_{1}$ and $F_{2}$ are usual isovector and isoscalar form factors which themselves are functions of the proton and neutron electric and magnetic form factors. These are highly constrained by electron scattering and we adopt the Kelly parameterization in this work~\cite{Kelly:2004hm}. The expression of the pseudo scalar form factor in terms of the axial one is the same of Eq.~\eqref{eq:PCAC}. The details on the different parameterizations adopted for $F_{A}$ are given in Secs.~\ref{sec:axialFF}. 
\vskip 0.12in
To treat amplitudes involving two-nucleon currents, the factorization ansatz of Eq.~\eqref{eq:fact:1b} can be generalized as 
\begin{align}
    |\psi_f^A \rangle \rightarrow |p p^\prime \rangle_a \otimes |\psi_f^{A-2}\rangle\, .
\end{align}
where $|p\,p^\prime\rangle_a=|p\,p^\prime\rangle-|p^\prime\,p\rangle$ is the anti-symmetrized state of two-plane waves with momentum $p$ and $p^\prime$. Following the work presented in Refs.~\cite{Benhar:2015ula,Rocco:2015cil,Rocco:2018mwt}, the pure two-body current component of the response tensor can be written as
\begin{align}
&R^{\mu\nu}_{\rm 2b}({\bf q},\omega)=\frac{V}{2} \int dE \frac{d^3k}{(2\pi)^3}  \frac{d^3k^\prime}{(2\pi)^3}\frac{d^3p}{(2\pi)^3}
\frac{m_N^4}{e({\bf k})e({\bf k^\prime})e({\bf p})e({\bf p^\prime})} \nonumber\\
 &\qquad \times  P_h^{\rm NM}({\bf k},{\bf k}^\prime,E) \sum_{ij}\, \langle k\, k^\prime | {j_{ij}^\mu}^\dagger |p\,p^\prime\rangle_a \langle p\,p^\prime |  j_{ij}^\nu | k\, k^\prime \rangle\nonumber\\
 &\qquad \times \delta(\omega-E+2m_N-e(\mathbf{p})-e(\mathbf{p}^\prime))\, .
\end{align}
In the above equation, the normalization volume for the nuclear wave functions $V=\rho / A$ with $\rho=3\pi^2 k_F^3/2$ depends on the Fermi momentum of the nucleus, which for $^{12}$C is taken to be $k_F=225$ MeV. 
 The factor $1/2$ accounts for the double counting that occurs in this notation (the product of the two direct terms is equal to the one of the two exchange terms). 
Differently from Refs.~\cite{Rocco:2018mwt}, the two-nucleon spectral function adopted in this work is not factorized into the product of two one-nucleon spectral functions, see Sec.~\ref{sub_sec:sf} for a detailed discussion. 

The two-body CC operator is given by the sum of four distinct interaction mechanisms, namely the pion in flight, seagull, pion-pole, and $\Delta$ excitations ~\cite{Hernandez:2007qq, Simo:2016ikv}
\begin{align}
j^\mu_{\rm CC}= (j^\mu_{\rm pif})_{\rm CC}+(j^\mu_{\rm sea})_{\rm CC}+ (j^\mu_{\rm pole})_{\rm CC}+ (j^\mu_{\Delta})_{\rm CC}.
\label{eq:jijCC}
\end{align}

\begin{figure}[h]
    \includegraphics[width=0.7\linewidth]{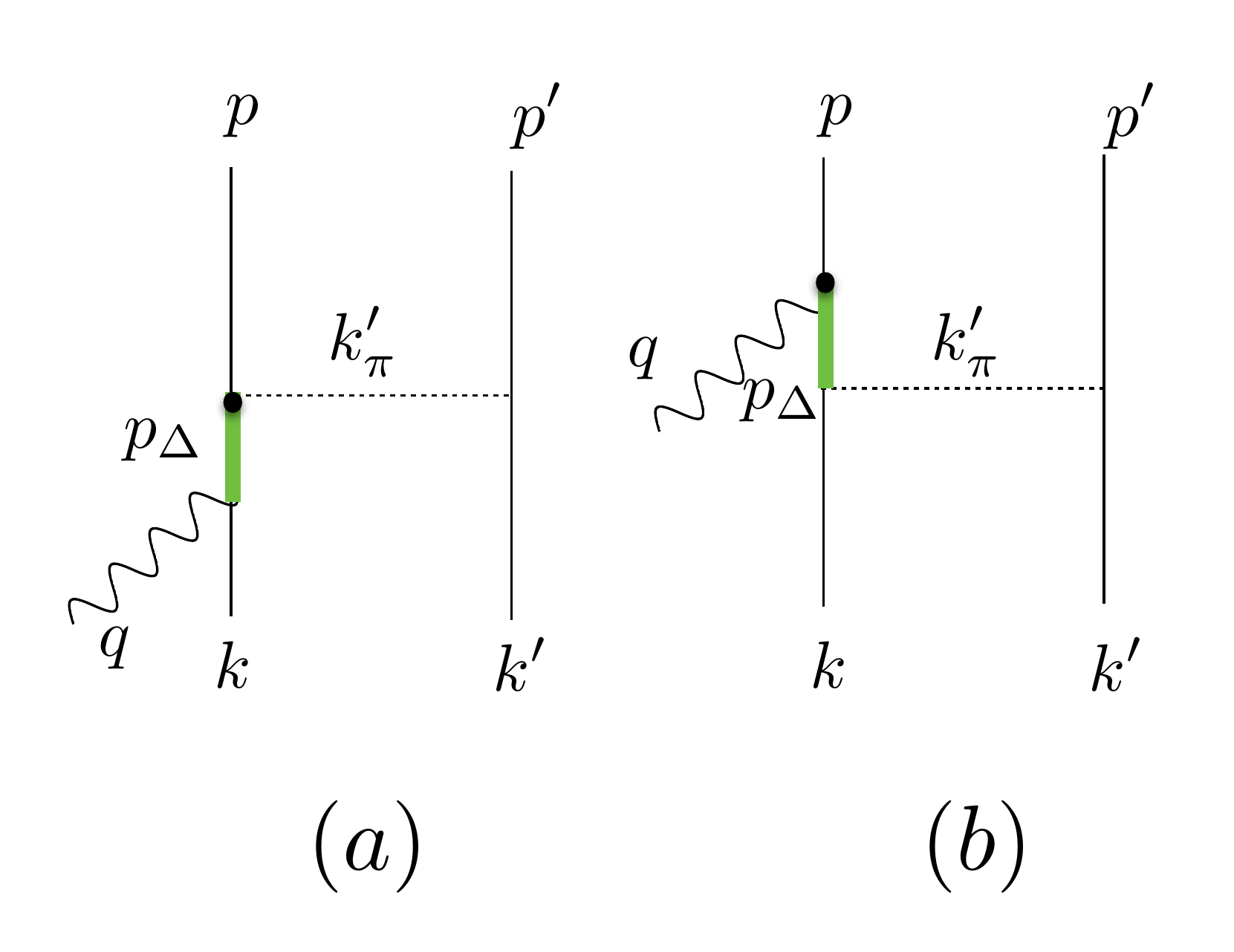} 
    \caption{Feynman diagrams describing of the first two contributions to the two-body currents associated with $\Delta$-excitation processes. Solid, thick green, and dashed lines correspond to nucleons, deltas, pions, respectively. The wavy line represents the vector boson.}
    \label{fig:j_Delta}
\end{figure}

Detailed expressions for the first four terms of Eq.~\eqref{eq:jijCC} can be found in~Refs.~\cite{Simo:2016ikv,Rocco:2018mwt}. Below, we only report the two-body current terms involving a $\Delta$-resonance in the intermediate state, as we find them to be the dominant contribution. Because of the purely transverse nature of this current, the form of its vector component is not subject to current-conservation constraints and its expression is largely model dependent, as discussed in the previous section. The current operator can be written as ~\cite{Hernandez:2007qq,Simo:2016ikv}:
\begin{align}
(j^\mu_\Delta)_{\rm CC}&=\frac{3}{2}\frac{f_{\pi NN} f^\ast}{m^2_\pi} \Bigg\{ 
\Bigg[ \Big( -\frac{2}{3}\tau^{(2)}_{\pm}+\frac{(\tau^{(1)}\times \tau^{(2)})_\pm}{3}\Big) \nonumber\\
&\times F_{\pi NN}(k^\prime_\pi) F_{\pi N \Delta} (k^\prime_\pi) (j^\mu_{a})_{(1)} \nonumber \\
&-\Big(\frac{2}{3}\tau^{(2)}_{\pm}+\frac{(\tau^{(1)}\times \tau^{(2)})_\pm}{3}\Big)_{\pm}\Big) \nonumber\\
&\times F_{\pi NN}(k^\prime_\pi) F_{\pi N \Delta} (k^\prime_\pi) (j^\mu_{b})_{(1)}\Bigg]\Pi(k^\prime_\pi)_{(2)}+(1\leftrightarrow 2)\Bigg\}
\label{delta:curr}
\end{align}
where $k^\prime$ and $p^\prime$ are the initial and final momentum of the second nucleon, respectively, while $k^\prime_\pi= p^\prime- k^\prime$ is the momentum of the $\pi$ exchanged in the two depicted diagrams of Fig.~\ref{fig:j_Delta}, $f^\ast$=2.14, and 
\begin{align}
&\Pi(k_\pi)=\frac{\gamma_5 \slashed{k}_\pi}{k_\pi^2-m^2_\pi}\ ,\\
&F_{\pi N \Delta}(k_\pi)=\frac{\Lambda^2_{\pi N\Delta}}{\Lambda^2_{\pi N\Delta}-k_\pi^2}\ , \\
&F_{\pi NN}(k_\pi)= \frac{\Lambda_\pi^2-m_\pi^2}{\Lambda_\pi^2-k_\pi^2}\label{fpinn}\ ,
\end{align}
with $\Lambda_{\pi N\Delta}=1150$ MeV and $\Lambda_{\pi}=1300$ MeV. The term $\tau_\pm= (\tau_x \pm i \tau_y)/2$ is the isospin raising/lowering operator.
In Eq.~\eqref{delta:curr}, $j^\mu_a$ and $j^\mu_b$ denote the $N\rightarrow \Delta$ transition vertices of diagram (a) and (b) of Fig.~\ref{fig:j_Delta}, respectively. The expression of $j^\mu_a$ is given by 
\begin{align}
j^\mu_a&=(j^\mu_a)_V+(j^\mu_a)_A\ ,\nonumber\\
(j^\mu_a)_V&=(k_\pi^\prime)^\alpha G_{\alpha\beta}(p_\Delta)\Big[\frac{C_3^V}{m_N}\Big(g^{\beta\mu}\slashed{q}-q^\beta\gamma^\mu\Big)\nonumber\\
&+\frac{C_4^V}{m_N^2}\Big(g^{\beta\mu}q\cdot p_\Delta-q^\beta p_\Delta^\mu\Big)\nonumber\\
&+\frac{C_5^V}{m_N^2}\Big(g^{\beta\mu}q\cdot k-q^\beta k^\mu + C_6^V g^{\beta\mu}\Big)
\Big]\gamma_5\ ,\nonumber\\
(j^\mu_a)_A&= (k_\pi^\prime)^\alpha G_{\alpha\beta}(p_\Delta)\Big[ \frac{C_3^A}{m_N}\Big(g^{\beta\mu}\slashed{q}-q^\beta\gamma^\mu\Big)\nonumber\\
&+\frac{C_4^A}{m_N^2}\Big(g^{\beta\mu}q\cdot p_\Delta-q^\beta p_\Delta^\mu\Big)\nonumber\\
&+ C_5^A g^{\beta\mu}+\frac{C_6^A}{m_N^2}q^\mu q^\alpha \Big],
\label{eq:delta:curr:ja}
\end{align}
 where $k$ is the momentum of the initial nucleon which absorbs the incoming momentum $\tilde{q}$ and $p_\Delta =\tilde{q}+k$, yielding $p^0_\Delta=e({\bf k})+\tilde{\omega}$. We introduced $\tilde{q}=(\tilde{\omega},{\bf q})$ to account for the fact that the initial nucleons are off-shell. A similar definition can be written down for $j^\mu_b$; more details are reported in Ref.~\cite{Rocco:2019gfb,Rocco:2018mwt}. 
For $C_3^V$ we adopted the model of Ref.~\cite{Lalakulich:2006sw}, yielding
\begin{align}
    C_3^V= \frac{2.13}{(1-q^2/M_V^2)^2}\frac{1}{1-q^2/(4 M_V^2)},
\end{align}
with $M_V=0.84$ GeV. Following the discussion of Ref.~\cite{Simo:2016ikv}, we neglected the terms $C_4^V$ and $C_5^V$ which are expected to be suppressed by $\mathcal{O}(k/m_N)$, while $C_6^V=0$ by conservation of the vector current. However, it is worth mentioning that including these terms in the current operator would not pose any conceptual difficulty. 
 To be consistent, in the axial part we only retain the leading contribution of Eq.~\eqref{eq:delta:curr:ja}, which is the term proportional to $C_5^A$ defined as ~\cite{Paschos:2003qr}
 \begin{align}
     C_5^A= \frac{1.2}{(1-q^2/M_{A\Delta})^2} \times \frac{1}{1-q^2/(3M_{A\Delta})^2)},
 \end{align}
 with $M_{A\Delta}= 1.05$ GeV. 

The Rarita-Schwinger propagator 
\begin{align}
G^{\alpha\beta}(p_\Delta)=\frac{P^{\alpha\beta}(p_\Delta)}{p^2_\Delta-M_\Delta^2},
\label{eq:free_delta}
\end{align}
is proportional to the spin 3/2 projection operator $P^{\alpha\beta}(p_\Delta)$.
In order to account for the possible decay of the $\Delta$ into a physical $\pi N$, we replace $M_\Delta \rightarrow M_\Delta - i \Gamma(p_\Delta)/2$~\cite{Dekker:1994yc,DePace:2003spn} where the last term is the energy dependent decay width given by
\begin{align}
\Gamma(p_\Delta)&=\frac{(4 f_{\pi N \Delta})^2}{12\pi m_\pi^2} \frac{|\mathbf{d}|^3}{\sqrt{s}} (m_N + E_d) R(\mathbf{r}^2)\, .
\label{eq:decay_width}
\end{align}
In the above equation, $(4 f_{\pi N \Delta})^2/(4\pi)=0.38$, $s=p_\Delta^2$ is the invariant mass, $\mathbf{d}$ is the decay three-momentum
in the $\pi N$ center of mass frame, such that
\begin{equation}
|\mathbf{d}|^2=\frac{1}{4s}[s-(m_N+m_\pi)^2][s-(m_N-m_\pi)^2]\,
\end{equation} 
and $E_d=\sqrt{m_N^2 + \mathbf{d}^2}$ is the associated energy. The additional factor
\begin{equation}
R(\mathbf{r}^2)=\left(\frac{\Lambda_R^2}{\Lambda_R^2-\mathbf{r}^2}\right),
\end{equation}
depending on the $\pi N$ three-momentum $\mathbf{r}$, with $\mathbf{r}^2=(E_d - \sqrt{m_\pi^2 + \mathbf{d}^2})^2-4\mathbf{d}^2$ and $\Lambda_R^2=0.95\, m_N^2$,
is introduced to improve the description of the experimental phase-shift $\delta_{33}$~\cite{Dekker:1994yc}.
The medium effects on the $\Delta$ propagator are accounted for by modifying the decay width as
\begin{align}
    \Gamma_\Delta(p_\Delta)\to \Gamma_\Delta(p_\Delta)-2\rm{Im}[U_\Delta(p_\Delta,\rho=\rho_0)],
\end{align}
where $U_\Delta$ is a density dependent potential obtained from a Bruckner-Hartree-Fock calculation using 
a coupled-channel $NN\oplus N\Delta\oplus \pi NN$ model~\cite{Lee:1983xu,Lee:1984us,Lee:1985jq,Lee:1987hd} and we fixed the density at the nuclear saturation value $\rho_0$ =0.16 fm$^3$. For a detailed analysis of medium effects in the MEC contribution for electron-nucleus scattering see Ref.~\cite{Rocco:2019gfb}.  


\subsection{Spectral Function}
\label{sub_sec:sf}

The spectral function of a nucleon with isospin $\tau_k=p,n$ and momentum ${\bf k}$ can be written as
\begin{align}
    P_{\tau_k}(\mathbf{k},E)&=\sum_n |\langle 0| [|k\rangle\, |\Psi_n^{A-1}\rangle]|^2 \delta(E+E_0-E_{n}^{A-1})\, .
\label{pke:hole}
\end{align}
where $|k\rangle$ is the single-nucleon state, $|0\rangle$ is the ground state of the Hamiltonian with energy $E_0$, while $|\Psi_n^{A-1}\rangle$ and $E_{n}^{A-1}$ are the energy eigenstates and eigenvalues of the remnant nucleus with $(A-1)$ particles.
The momentum distribution of the initial nucleon is obtained by integrating the spectral function over the removal energy  
\begin{equation}
n_{\tau_k}(k) = \int dE P_{\tau_k}(\mathbf{k},E)\,.
\end{equation}
We can rewrite the spectral function as a sum of a mean field (MF) and a correlation (corr) term. 
For simplicity, let us focus on the $\tau_k=p$ case; the MF contribution of the proton spectral function is obtained by 
considering only bound $A-1$ states of the remnant nucleus
\begin{align}
P_p^{\rm MF} (\mathbf{k},E)&=\sum_n| \langle 0^{A}| [|k\rangle \otimes |\Psi_n^{A-1}\rangle]|^2 \nonumber\\
&\times \delta\Big(E-B^{A}_0+B_n^{A-1}-\frac{k^2}{2m^{A-1}_n}\Big)\, ,
\end{align}
where $B^{A}_0$ and $B_n^{A-1}$ are the binding energies of the initial and the bound $A-1$ spectator nucleus with mass and $m_n^{A-1}$. The momentum-space overlaps $\langle 0^{A} | [|k\rangle \otimes |\Psi_n^{A-1}\rangle$ pertaining to the p-shell contributions are computed by Fourier transforming the variational Monte Carlo (VMC) radial overlaps for the transitions ~\cite{overlaps_web}: 
\begin{align*}
    ^{12}{\rm C}(0^+)&\rightarrow ^{11}{\rm B}(3/2^-)+p \nonumber\\
    ^{12}{\rm C}(0^+)&\rightarrow ^{11}{\rm B}(1/2^-)+p \nonumber\\
    ^{12}{\rm C}(0^+)&\rightarrow ^{11}{\rm B}(3/2^-)^\ast+p\, .
\end{align*}
The calculation of the s-shell mean-field contribution involves non trivial difficulties for the VMC method, as it would require to evaluate  the spectroscopic overlaps for the transitions to all the possible excited states of $^{11}{\rm B}$ with $J^P=(1/2^+)$. To overcome this limitation, we used the VMC overlap associated with the $^{4}{\rm He}(0^+)\rightarrow {}^{3}{\rm H}(1/2^+)+p$ transition and applied minimal changes to the quenching factor which is needed to reproduce the integral of the momentum distribution up to $k_F= 1.15$ fm$^{-1}$. More details about the adopted procedure are discussed in Ref.~\cite{CLAS:2022wvz}.  

The correlation contribution to the SF is given by
\begin{align}
P_p^{\rm corr}&(\mathbf{k},E)=\sum_n \int \frac{d^3 k^\prime}{(2\pi)^3} |\langle 0^A| [|k\rangle\,  |k^\prime\rangle \, |\Psi_n^{A-2}\rangle]|^2\nonumber\\
&\times \delta(E+E_0-e(\mathbf{k}^\prime)-E_{n}^{A-2})\, \nonumber\\
&=  \mathcal{N}_p \sum_{\tau_{k^\prime}=p,n} \int \frac{d^3 k^\prime}{(2\pi)^3} \Big[ n_{p,\tau_{k^\prime}}(\mathbf{k},\mathbf{k}^\prime)\nonumber \\
&\qquad \times \delta \Big(E-B_0-e(\mathbf{k}^\prime)+\bar{B}_{A-2}-
\frac{(\mathbf{k}+\mathbf{k}^\prime)^2}{2m_{A-2}}\Big)\Big]\, , 
\label{pke:2b}
\end{align}
to derive the last expression, we used a completeness relation and assumed that the $(A -2)$-nucleon binding energy is narrowly distributed around a central value $\bar{B}_{A-2}$. The mass of the recoiling $A-2$ system is denoted by $m_{A-2}$ and $\mathcal{N}_p$ is an appropriate normalization factor. 
We started from the VMC two-nucleon momentum distribution $n_{\tau_k,\tau_{k^\prime}}(\mathbf{k},\mathbf{k}^\prime)$ of Ref.~\cite{nkk_web}, but in order to isolate the contribution of short-range correlated nucleons we performed cuts in the relative momentum of the pairs, requiring that the overall normalization and shape of the one-nucleon momentum distributions are correctly recovered.  

In this work, we only consider the MF contribution to the two-nucleon spectral function, e.g. we neglect contributions where more than two nucleons are emitted. We write $P_{\tau_k,\tau_k^\prime}^{\rm MF}$ as
\begin{align}
P_{\tau_k,\tau_k^\prime}^{\rm MF}(\mathbf{k},\mathbf{k}^\prime,E)&=
n_{\tau_k,\tau_{k^\prime}}(\mathbf{k},\mathbf{k}^\prime)\nonumber \\
& \times \delta \Big(E-B_0+\bar{B}_{A-2}-
\frac{\mathbf{K}^2}{2m_{A-2}}\Big)\, . 
\end{align}
where $\mathbf{K}=\mathbf{k}+\mathbf{k}^\prime$ is the total momentum of the pair. 
It has been argued that the strong isospin dependence of the two-nucleon momentum distribution, supported by experimental data, persists for nuclei even larger than $^{12}$C~\cite{Wiringa:2013ala,Weiss:2015mba,Weiss:2018tbu,Lonardoni:2018sqo}. In Refs.~\cite{Rocco:2018mwt,Rocco:2019gfb} $P(\mathbf{k},\mathbf{k}^\prime,E)$ has been written as the product of two one-nucleon spectral functions assuming that long-range correlations are absent. In the present work, we go beyond this approximation and adopt a novel definition of the two-nucleon spectral function that retains correlations between the two struck particles. A detailed comparison of the results obtained with these different parametrizations of the two-nucleon spectral functions will be carried out in a future work and validated against electron scattering data.

\section{Nucleon axial form factor effects and uncertainties}\label{sec:axialFF}

Imprecise knowledge of the nucleon axial form factor, $F_{A}$, might lead to significant systematic uncertainties in neutrino-nucleus cross-section calculations.
This is in contrast with the nucleon vector form factors in Eq.~\eqref{eq:FF}, which can be precisely determined using high-statistics electron scattering experiments~\cite{Leitner:2008ue,JeffersonLabHallA:2018zyx,Barbieri:2019ual,Ankowski:2020qbe,CLAS:2021neh}.
Current experimental constraints on nucleon axial form factors come from beta decay measurements, neutrino scattering on nuclear targets, and pion electroproduction~\cite{Mann:1973pr,Barish:1977qk,Baker:1981su,Miller:1982qi,Kitagaki:1983px,Radecky:1981fn,Kitagaki:1986ct,Kitagaki:1990vs,Bhattacharya:2011ah,Meyer:2016oeg}. These give weak constraints on $F_{A}$ in comparison to the vector form factors, as beta decay is only sensitive the absolute normalization $g_A = F_A(0)$ and neutrino scattering and pion electroproduction experiments are limited by both statistics and nuclear modeling uncertainties.

As LQCD calculations of nucleon form factors mature~\cite{Shintani:2018ozy,Jang:2019vkm,RQCD:2019jai,Alexandrou:2020okk,Ishikawa:2021eut,Meyer:2021vfq,Park:2021ypf,Djukanovic:2022wru}, it becomes increasingly important to quantify the level of axial form factor precision required to achieve a given level of neutrino-nucleus cross-section accuracy.
This is challenging because axial form factor effects on flux-averaged neutrino-nucleus cross sections can be difficult to disentangle from nuclear effects such as two-body currents, as evident for example in the differences between theoretical descriptions of MiniBooNE data with either an unexpectedly slow falloff of the axial form factor with increasing momentum transfer~\cite{MiniBooNE:2010bsu} or with larger than anticipated contributions from two-body current effects~\cite{Martini:2010ex,Martini:2011wp,Nieves:2011pp,Nieves:2011yp,Megias:2014qva,Gonzalez-Jimenez:2014eqa,Pandey:2014tza}.
This ambiguity between one- and two-body current effects on flux-averaged cross sections makes it essential to quantify the role of the nucleon axial form factor in neutrino-nucleus cross-section calculations using nuclear effective theories that provide a consistent theoretical decomposition between one- and two-body current contributions.
The remainder of this section discusses how to quantify nucleon axial form factor effects on neutrino-nucleus cross-section calculations based on the model-independent $z$ expansion and how to estimate nucleon axial form faction precision needs using the GFMC and spectral function methods discussed above.

\subsection{Parametrization}
\label{sub_sec:fa_param}

Historically a dipole parametrization has often been used for the axial form factor
\begin{equation}
    F_{A}(Q^{2}) = \frac{g_{A}}{(1+Q^{2}/M_{A}^{2})^2},
\end{equation}
where $g_{A} = 1.2723(23)$ has been measured from neutron beta decay~\cite{ParticleDataGroup:2016lqr}, and $M_{A} = 1.014 \pm 0.014$ GeV~\cite{Bodek:2007ym}. 
However, this one-parameter form is not expressive enough to describe the shape of the axial form factor predicted by QCD.
It has been demonstrated in Refs.~\cite{Bhattacharya:2011ah,Meyer:2016oeg} that assuming that the dipole parameterization is valid when fitting to experimental results can lead to form factor fits with uncertainties that are underestimated by a factor of $\sim 5$ in comparison to those determined using fits based off a model-independent $z$ expansion.
\vskip 0.12in
Axial form factors in QCD are analytic functions of $Q^2$ for $Q^2 = -t > -t_c$, where $t_c$ is the location of the $t$-channel cut, which enables an analytic function $z(Q^2)$ to be defined as~\cite{Hill:2006ub,Hill:2010yb,Bhattacharya:2011ah}
\begin{equation}
  z(Q^2) = \frac{\sqrt{t_c + Q^2} - \sqrt{t_c - t_0}}{\sqrt{t_c + Q^2} + \sqrt{t_c - t_0}},
  \label{eq:z}
\end{equation}
where $t_0$ is an arbitrary parameter whose choice is discussed in Sec.~\ref{sec:axial_ff_results} below.
For $F_A$ the cut starts at $t_c = 9m_\pi^2$.
Because $|z| < 1$, the axial form factor can be expanded as a power series in $z(Q^2)$ for the $Q^2 > 0$ domain of interest for neutrino-nucleus scattering,
\begin{equation}
  F_{A}(Q^2) = \sum_{k=0}^{\infty} a_{k} \, z(Q^2)^{k} \approx \sum_{k=0}^{k_{\rm max}} a_{k} \, z(Q^2)^{k},
  \label{eq:zexp}
\end{equation}
where the $z$ expansion coefficients $a_k$ include nucleon structure information and $k_{\rm max}$ is a truncation parameter required to make the number of expansion parameters finite.
The parameter $a_0$ can be fixed by the sum rule
\begin{equation}
  \sum_{k=0}^{k_{\rm max}} a_k z(0)^k = g_A. \label{eq:sum0}
\end{equation}
Constraints on the $a_k$ are also obtained by enforcing the correct large $Q^2$ behavior of the axial form factor, which is predicted by perturbative QCD to be $Q^{-4}$ up to logarithmic corrections~\cite{Lepage:1980fj}.
This asymptotic $Q^{-4}$ behavior can be enforced by demanding that $F_A(Q^2)$ and its first three derivatives with respect to $1/Q$ vanish for asymptotically large $Q^2$, corresponding to $z=1$, which is equivalent to
\begin{equation} \label{eq:2}
\left. \frac{d^n}{dz^n} F_A \right|_{z=1} = 0 \; \; \; ; \; \; \; n = 0,1,2,3,
\end{equation}
and therefore leads to the sum rules~\cite{Lee:2015jqa}
\begin{equation} \label{eq:sum_constraints}
  \sum_{k=n}^{k_{\rm max}} \frac{k!}{(k-n)!} \  a_k = 0 \; \; \; ; \; \; \; n = 0,1,2,3 .
\end{equation}
In practice, these constraints can be satisfied by first determining the $k_{\rm max}$ and $a_1,\ldots,a_{k_{\rm max}}$ preferred by a fit to data (with $a_0$ either treated as an additional independent parameters or as being fixed by the constraint Eq.~\eqref{eq:sum0}), 
and then replacing $k_{\rm max}$ with $k_{\rm max}+4$ and solving for the four unconstrained coefficients using Eq.~\eqref{eq:sum_constraints}.
The remaining $a_k$ must then be fixed by information on the $Q^2$-dependence of the axial form factor
determined theoretically using LQCD calculations or  experimentally by fitting neutrino-nucleus scattering and/or pion electroproduction data.
\vskip 0.12in
The $z$ expansion can be used to provide a model-independent definition of the dependence of neutrino-nucleus cross-section uncertainties on nucleon axial form factor uncertainties.
Any function $\sigma(F_A,X)$ that depends on the axial form factor, as well as any number of additional independent form factors and parameters schematically denoted by $X$, can be interpreted as a function of the axial form factor $z$ expansion coefficients, $\sigma(a_k,X)$.
Neglecting correlations among the $a_k$ and between $X$ and the $a_k$, the relative uncertainties of $\sigma$, $a_k$, and $X$ can be expressed as
\begin{equation}
  \begin{split}
    \frac{\delta \sigma}{|\sigma|} &= \left| \frac{X}{\sigma} \right| \frac{\delta X}{|X|} + \sum_k \left| \frac{\partial \sigma}{\partial a_k} \right| \left| \frac{a_k}{\sigma} \right| \frac{\delta a_k}{|a_k|} \\
    & \geq \left| \frac{\partial \sigma}{\partial a_k} \right| \left| \frac{a_k}{\sigma} \right| \frac{\delta a_k}{|a_k|}, \label{eq:chain_rule}
  \end{split}
\end{equation}
where the inequality holds for each $a_k$ individually by positivity of each term in the sum.
In particular, Eq.~\eqref{eq:chain_rule} can be applied to neutrino-nucleus differential cross sections $\frac{d\sigma}{dT_\mu d\cos\theta_\mu}$ at the values of $T_\mu$ that maximize the cross-section for a given $\cos\theta_\mu$, denoted $d\sigma_P \equiv \left.\frac{d\sigma}{dT_\mu d\cos\theta_\mu}\right|_{\rm peak}$, to give
\begin{equation}
  \begin{split}
    \frac{\delta \left( d\sigma_P \right)}{d\sigma_P} & \geq D_k(\theta_\mu)  \frac{\delta a_k}{|a_k|}, \label{eq:chain_rule_dsigma}
  \end{split}
\end{equation}
where
\begin{equation}
  D_k(\theta_\mu) =  \left| \frac{\partial d\sigma_P}{\partial a_k} \right|  \frac{\left|a_k\right|}{d\sigma_P}. \label{eq:chain_rule_coefficient}
\end{equation}
A simple way to approximate the derivative appearing in $D_k(\theta)$ is to compute the finite difference,
\begin{equation}
  \begin{split}
    \frac{\partial d\sigma_P}{\partial a_k} \approx& \frac{1}{\delta a_k}\left[ d\sigma_P(a_k + \delta a_k)  - d\sigma_P(a_k) \right], \label{eq:finite_diff} 
  \end{split}
\end{equation}
where $\delta a_k$ is a small parameter and $d\sigma_P(a_k + \delta a_k)$ denotes the peak differential cross section calculated with $a_k$ replaced by $a_k + \delta a_k$ and $a_i$ held fixed for $i\neq k$.
The accuracy of the finite-difference approximation can be validated by performing calculations with different values of $\delta a_k$ and verifying the stability of the resulting derivative estimate.

Since neutrino mixing angle determinations at oscillation experiments are sensitive to the relative heights of the cross-section peaks at the near and far detectors, the relative uncertainty in $\left.\frac{d\sigma}{dT_\mu d\cos\theta_\mu}\right|_{\rm peak}$ provides a reasonable measure of the cross-section uncertainty entering neutrino oscillation analyses.\footnote{Further information about the uncertainty needs of neutrino oscillation experiments could be obtained by applying Eq.~\eqref{eq:chain_rule} directly to the neutrino mass and mixing parameters derived from an oscillation analysis with for example $\frac{\partial \delta_{CP}}{\partial a_k}$ estimated in analogy to Eq.~\eqref{eq:finite_diff}. This requires a detailed analysis including for example neutrino energy reconstruction and detector efficiency effects and is left to future work. }
The relative uncertainty on $a_k$ required for axial form factor uncertainties to not saturate a given cross-section uncertainty budget according to this measure is then given by the desired cross-section uncertainty times $1/D_k(\theta)$.

\subsection{Form Factor Results}\label{sec:axial_ff_results}

Neutrino scattering experiments typically involve nuclear targets, and the challenges of determining nucleon axial form factors from data arise both from scarcity of precise experimental data and from the need to understand nuclear effects that are present in data. Relatively theoretically clean determinations can be made using neutrino-deuteron scattering data obtained from bubble-chamber experiments at ANL, BNL, and FNAL in the 70s and 80s ~\cite{Mann:1973pr,Barish:1977qk,Baker:1981su,Miller:1982qi,Kitagaki:1983px}, which were recently analyzed in Ref.~\cite{Meyer:2016oeg} using the $z$ expansion to parameterize the nucleon axial form factor. It was found that uncertainties on the axial form factor shape were significantly underestimated by previous analysis that assumed an over-constraining dipole form factor shape, and in particular the uncertainties on the axial radius are estimated to be five times larger using fits to the $z$ expansion than using dipole fits. The relation between the free nucleon and deuteron cross sections need for this analysis is parameterized as a $Q^2$-dependent but $E_\nu$-independent function taken from nuclear theory calculations~\cite{Singh:1971md,Shen:2012xz}.
A 10\% systematic uncertainty is included in the analysis of Ref.~\cite{Meyer:2016oeg}  to account for nuclear model uncertainties and other systematic corrections.

\begin{figure}[t]
    \includegraphics[width=\linewidth]{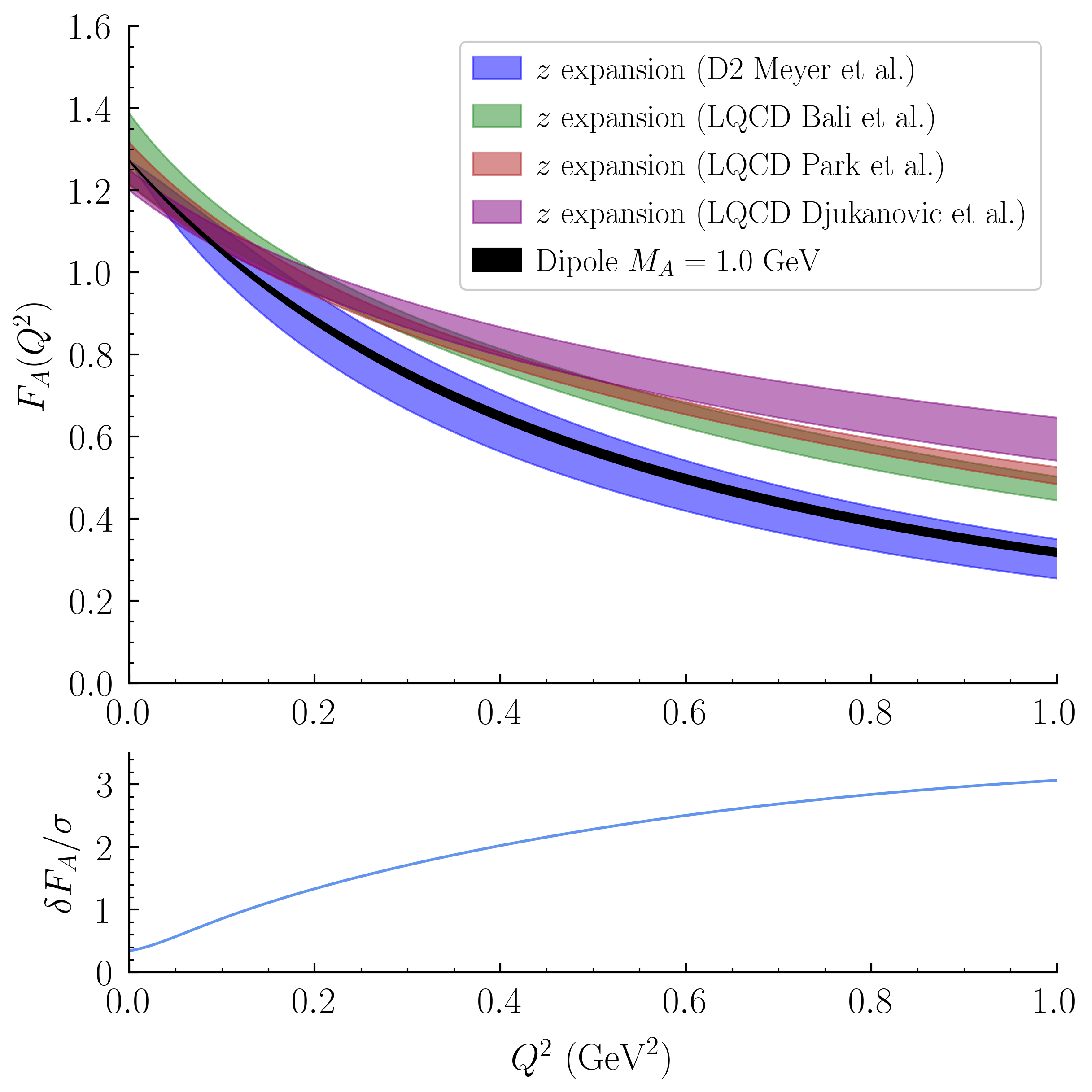} 
    \caption{The nucleon axial form factor $F_{A}(Q^2)$ determined using fits to neutrino-deuteron scattering data using the model-independent $z$ expansion from Ref.~\cite{Meyer:2016oeg} (D2 Meyer et al.) are shown as a blue band in the top panel. LQCD results are shown for comparison from Ref.~\cite{RQCD:2019jai} (LQCD Bali et al., green), Ref.~\cite{Park:2021ypf} (LQCD Park et al., red) and Ref.~\cite{Djukanovic:2022wru} (LQCD Djukanovic et al., purple). Bands show combined statistical and systematic uncertainties in all cases, see the main text for more details. A dipole parameterization with $M_A = 1.0$ GeV and a $1.4\%$ uncertainty~\cite{Bodek:2007ym} is also shown for comparison (black). The lower panel shows the absolute value of the difference between D2 Meyer et al. and LQCD Bali et al. results divided by their uncertainties added in quadrature, denoted $\delta F_{A}/\sigma$; very similar results are obtained using the other LQCD results.}
    \label{fig:FF}
\end{figure}

Alternative strategies for constraining nucleon axial form factors without nuclear uncertainties are to perform neutrino scattering experiments with hydrogen targets, see Ref.~\cite{Alvarez-Ruso:2022ctb} for a discussion of current proposals and challenges, and to calculate nucleon axial form factors theoretically using LQCD, as reviewed in Refs.~\cite{Kronfeld:2019nfb,Meyer:2022mix,Ruso:2022qes}.
There is a long history of LQCD studies of nucleon form factors
~\cite{Liu:1992ab,Alexandrou:2007eyf} including recent and ongoing work by multiple groups~\cite{Shintani:2018ozy,Jang:2019vkm,Jang:2019jkn,RQCD:2019jai,Alexandrou:2020okk,Ishikawa:2021eut,Meyer:2021vfq,Park:2021ypf,Djukanovic:2022wru}.
Challenging systematic uncertainties arising from $N\pi$ excited-state contamination have been identified in axial form factor calculations~\cite{Bar:2018xyi,Bar:2019gfx,Jang:2019vkm,RQCD:2019jai,Park:2021ypf,Barca:2021iak}, and recent calculations explicitly accounting for these effects have shown consistency with identities following from axial ward identities and the assumption of negligible excited-state contamination~\cite{Jang:2019vkm,RQCD:2019jai,Park:2021ypf}.
Although further study of these excited-state effects is needed~\cite{Barca:2021iak}, calculations have now been performed using approximately physical quark masses and continuum, infinite-volume extrapolations~\cite{RQCD:2019jai,Park:2021ypf,Djukanovic:2022wru}.
In each of these LQCD calculations, axial form factors are parameterized using the $z$ expansion and fit to nucleon axial-current matrix elements at several discrete values of $Q^2$, though different strategies for parameterizing discretization effects and other sources of systematic uncertainty are employed.\footnote{The uncertainties reported in Ref.~\cite{Park:2021ypf} are inflated by a factor of 3 to account for systematic uncertainties arising from lattice spacing and quark mass effects~\cite{Alvarez-Ruso:2022ctb}.}

The LQCD nucleon axial form factor results of Refs.~\cite{RQCD:2019jai,Park:2021ypf,Djukanovic:2022wru} are shown in Fig.~\ref{fig:FF} along with the form factor results determined from experimental neutrino-deuteron scattering data in Ref.~\cite{Meyer:2016oeg}.
Fits were performed using results with $Q^2 \leq 1$ GeV$^2$ in Refs.~\cite{Meyer:2016oeg,RQCD:2019jai,Park:2021ypf} and with $Q^2 \leq 0.7$ GeV$^2$ in Ref.~\cite{Djukanovic:2022wru} with the parameterization provided by the $z$ expansion used to extrapolate form factor results to larger $Q^2$. 
Clear agreement between different LQCD calculations can be seen. However, the LQCD axial form factor results are 2-3$\sigma$ larger than the results of Ref.~\cite{Meyer:2016oeg} for $Q^2 \gtrsim 0.3\text{ GeV}^2$. The effects of this form factor tension on neutrino-nucleus cross section predictions is studied using nuclear many-body calculations with the GFMC and SF methods in Sec.~\ref{sec:results} below. 
The LQCD results of Refs.~\cite{RQCD:2019jai,Park:2021ypf} lead to nearly indistinguishable cross-section results that will be denoted ``LQCD Bali et al./Park et al.'' or ``LQCD'' below and used for comparison with the deuterium bubble-chamber analysis of Ref.~\cite{Meyer:2016oeg}, denoted ``D2 Meyer et al.'' or ``D2'' below.

\begin{figure*}
    \includegraphics[width=\textwidth]{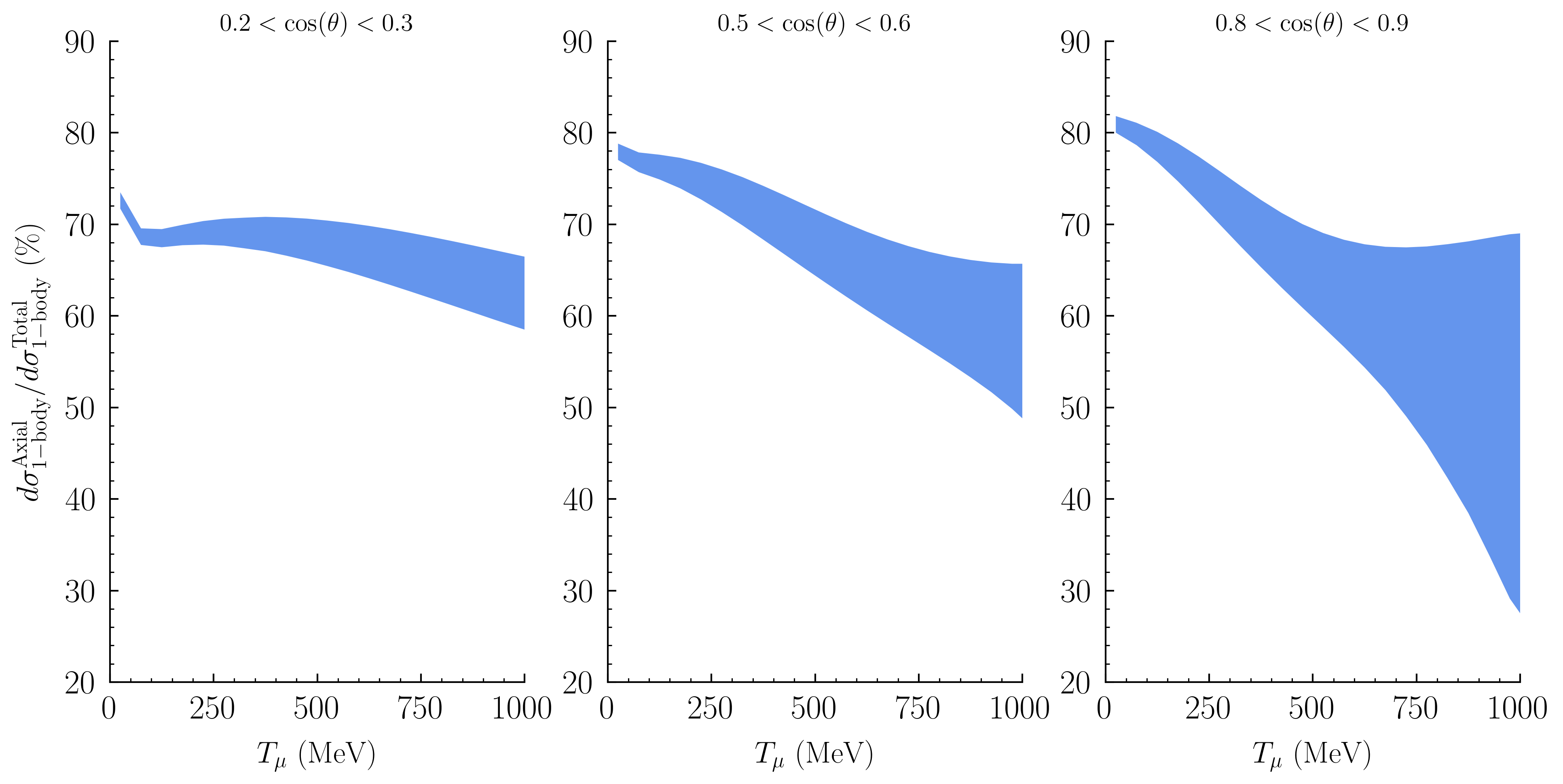} 
    \caption{The fractional contribution of the axial form factor to the one-body cross section for MiniBooNE kinematics in selected bins of $\cos\theta_{\mu}$ predicted using the SF formalism shown in Fig.~\ref{fig:datavstheory_sep_MB}. The central value and uncertainty of the axial form factor correspond to the D2 Meyer et al. $z$ expansion results~\cite{Meyer:2016oeg}.
Figures are cut off at $T_{\mu} = 1$ GeV because the cross sections become extremely small beyond this point and relative cross-section uncertainties become large.}
    \label{fig:axial_cont}
\end{figure*}

\section{Flux-Averaged Cross Section Results}
\label{sec:results}

To evaluate both the nuclear model and nucleon axial form factor dependence of neutrino-nucleus cross-section predictions and their agreement with data, the GFMC and spectral function methods are used to predict flux-averaged cross sections that can be compared with data from the T2K and MiniBooNE experiments. The MiniBooNE data for this comparison is a double differential CCQE measurement where the main CC1$\pi$+ background has been subtracted using a tuned model~\cite{MiniBooNE:2010bsu}, and the T2K data is a double differential CC0$\pi$ measurement~\cite{T2K:2016jor}.
Muon neutrino flux-averaged cross sections were calculated from
\begin{equation}
    \frac{d\sigma}{dT_{\mu}d\cos\theta_{\mu}} = \int dE_{\nu}\phi(E_{\nu})\frac{d\sigma (E_{\nu})}{dT_{\mu}d\cos\theta_{\mu} },
\end{equation}
where $\phi(E_{\nu})$ are the normalized $\nu_{\mu}$ fluxes from MiniBooNE and T2K. Details on the neutrino fluxes for each experiment can be found in the references above. $\frac{d\sigma (E_{\nu})}{dT_{\mu}d\cos\theta_{\mu} }$ are the corresponding inclusive cross sections computed using the GFMC and SF methods as described in Sec.~\ref{sec:methods}.

The fractional contribution of the axial form factor to the one-body piece of the MiniBooNE flux-averaged cross section is determined by including only pure axial and axial-vector interference terms in the cross section and shown in Fig.~\ref{fig:axial_cont}.
These pure axial and axial-vector interference terms account for half or more of the total one-body cross section for most $T_\mu$ and $\cos\theta_{\mu}$, which emphasizes the need for an accurate determination of the nucleon axial form factor.

\begin{figure*}[t]
    \centering
    \includegraphics[width=\textwidth]{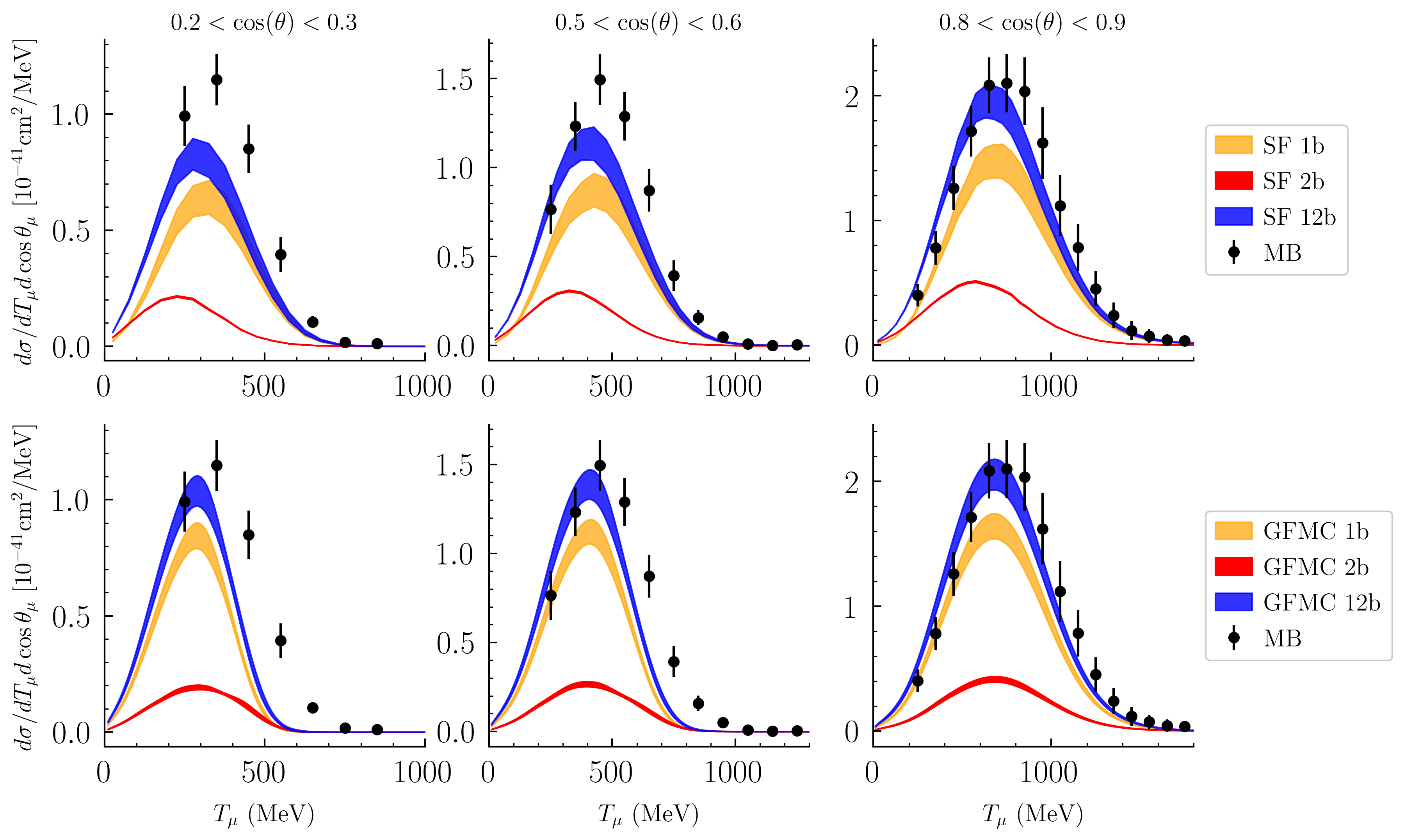}
    \caption{Breakdown into one- and two-body current contributions of the $\nu_{\mu}$ flux-averaged differential cross sections for MiniBooNE: 1b and 2b denotes one- and two-body current contributions while 12b denotes the total sum of these contributions. The top panel shows Spectral Function predictions in three bins of $\cos\theta_{\mu}$ with the one-body contributions in orange, two-body contributions in red, and the total in blue. The lower panel shows GFMC predictions with the same breakdown between one- and two-body current contributions, although the two-body results include interference effects only in the GFMC case.  The D2 Meyer et al. $z$ expansion results for $F_{A}$ are used in both cases~\cite{Meyer:2016oeg}. }
    \label{fig:datavstheory_sep_MB}
\end{figure*}

\begin{figure*}[t]
    \centering
    \includegraphics[width=\textwidth]{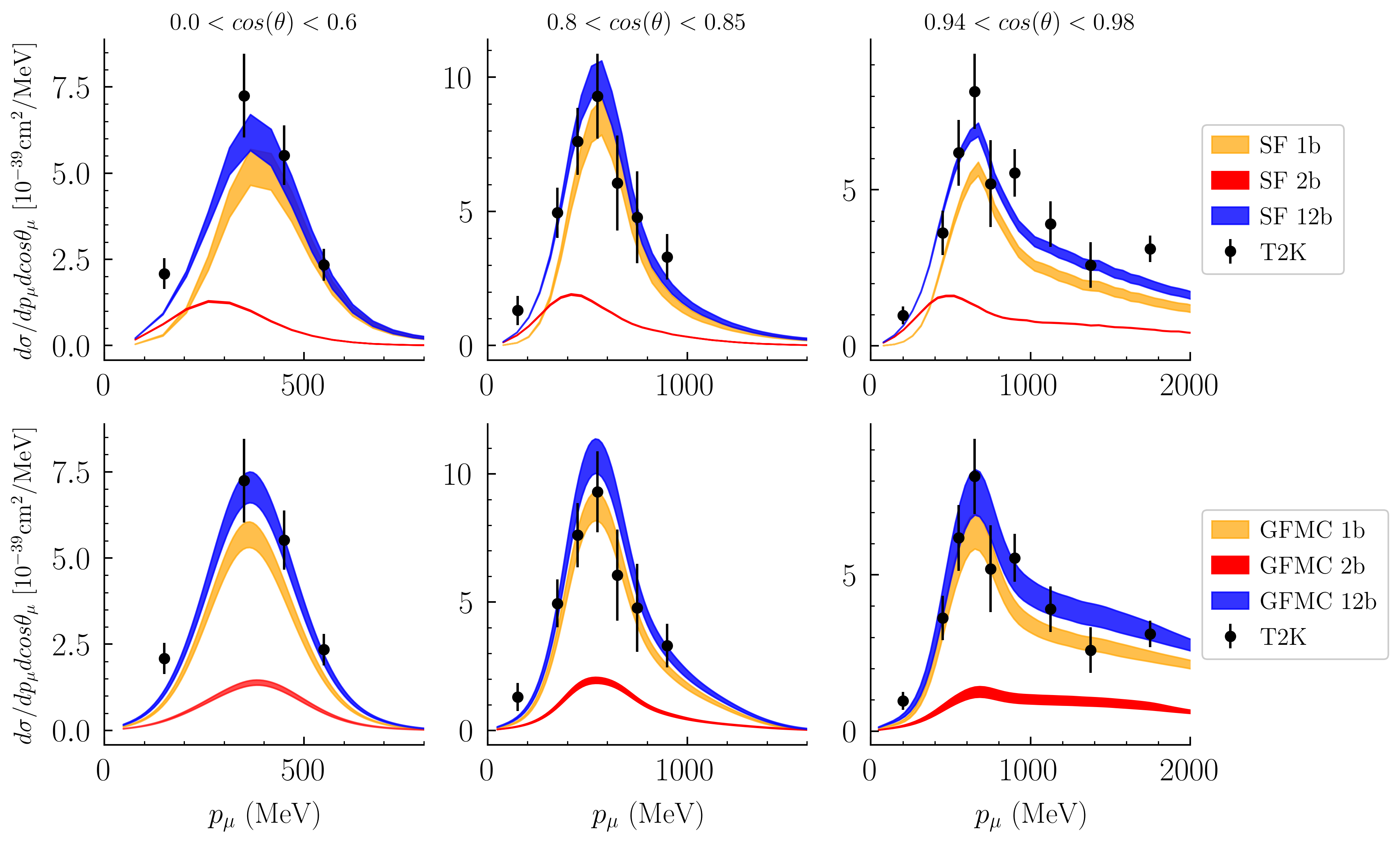}
    \caption{Breakdown into one- and two-body current contributions of the $\nu_{\mu}$ flux-averaged differential cross sections for T2K. The color code is as in Fig.~\ref{fig:datavstheory_sep_MB}.}
    \label{fig:datavstheory_sep_T2K}
\end{figure*}

Figures~\ref{fig:datavstheory_sep_MB} and~\ref{fig:datavstheory_sep_T2K} show the GFMC and SF predictions for MiniBooNE and T2K, respectively, including the breakdown into one-body and two-body contributions. For these comparisons we use the D2 Meyer et al. $z$ expansion for $F_{A}$. Two features of the calculations should be noted before discussing the results of these comparisons.
First, the uncertainty bands in the SF come only from the axial form factor, while the GFMC error bands include axial form factor uncertainties as well as a combination of GFMC statistical errors and uncertainties
associated with the maximum-entropy inversion. Secondly, the axial form factor enters into the SF only in the one-body term, in contrast to the GFMC prediction where it enters into both the one-body and one and two-body interference term.

Below in Table~\ref{table:quant} we quantify the differences between GFMC and SF predictions for both MiniBooNE and T2K. The percent difference in the differential cross sections at each model's peak are shown. The GFMC predictions are up to $20\%$ larger in backwards angle regions for MiniBooNE and $13\%$ larger for T2K in the same backward region. The agreement between GFMC and SF predictions is better at more forward angles but a $5$-$10\%$ difference persists. 

Figure ~\ref{fig:datavstheory_sep_MB} shows relatively good agreement between SF predictions and MiniBooNE cross-section results for a range of kinematics, with SF underestimating data in the lowest bins of $\cos\theta_{\mu}$ by up to $20\%$ at the peak. The location of the peak is correctly predicted, and the agreement increases at forward angles. GFMC predictions for MiniBooNE see a similar deficit of events at low $\cos\theta_{\mu}$. The enhanced disagreement for GFMC at small $\cos\theta_{\mu}$ and large $T_{\mu}$ is expected as these bins occur at larger $Q^2$ where relativistic effects become important. 
Comparisons with T2K for both SF and GFMC in Fig.~\ref{fig:datavstheory_sep_T2K} are in excellent agreement with the data even for backwards angles. The GFMC prediction here is expected to be more accurate as the T2K $\nu_{\mu}$ flux has a smaller tail at high neutrino energy. It is interesting to note the shift in the location of the two-body peak between GFMC and SF with both MiniBooNE and T2K fluxes, the GFMC two-body contribution peaking in roughly the same location as the one-body, whereas the SF two-body peak is shifted to lower $T_{\mu}$. This is again because the GFMC two-body curve is dominated by the one- and two-body current interference which roughly peaks in the same location as the pure one-body term. In the SF calculation, the one- and two-body current interference term yielding two nucleon emission has been studied in Refs.~\cite{Rocco:2015cil,Benhar:2015ula} and found to be negligible. Therefore, in this work only the pure two-body contribution has been included which requires large energy transfer and thus peaks at lower $T_{\mu}$. Accounting for one- and two-body current interference with a single nucleon emitted in the final state involves some developments in the SF formalism and it will be the subject of a future work. 

\begin{table*}[t]
\centering
 \begin{ruledtabular}
 \begin{tabular}{cccc} 
   MiniBooNE & $0.2 < \cos\theta_{\mu} < 0.3$ & $0.5 < \cos\theta_{\mu} < 0.6$ & $0.8 < \cos\theta_{\mu} < 0.9$ \\\hline

GFMC/SF difference in $d\sigma_{\rm{peak}}$ $(\%)$  & $22.8$ & $20.3$ & $5.6$ \\
 \end{tabular}
   \vspace{10pt}
 \begin{tabular}{cccc} 
 T2K & $0.0 < \cos\theta_{\mu} < 0.6$ & $0.80 < \cos\theta_{\mu} < 0.85$ & $0.94 < \cos\theta_{\mu} < 0.98$ \\\hline

GFMC/SF difference in $d\sigma_{\rm{peak}}$ $(\%)$ & $13.4$ & $7.3$ & $10.0$ \\
 \end{tabular}
 \end{ruledtabular}
\caption{Difference in value of $\frac{d\sigma (E_{\nu})}{dT_{\mu}d\cos\theta_{\mu} }$ at the quasielastic peak computed using GFMC and SF methods for MiniBooNE and T2K flux-averaged double-differential cross sections. \label{table:quant} }
\end{table*}

\begin{figure*}[t]
    \centering
    \includegraphics[width=0.8\textwidth]{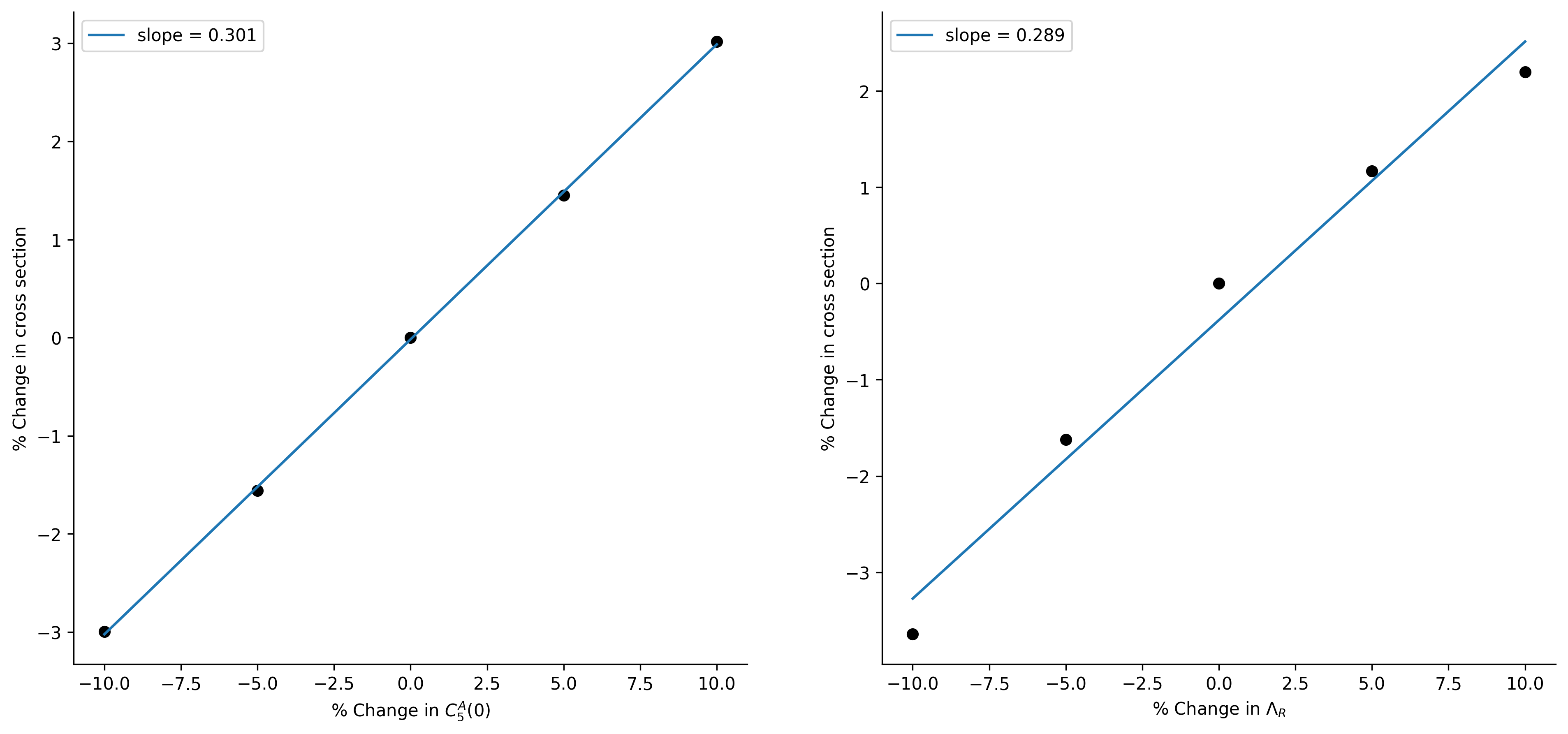}
    \caption{Percent change in the value of the MiniBooNE flux-averaged cross section for $0.5 < \cos\theta_{\mu} < 0.6$ vs. percent change in two parameters describing $\Delta$ resonance production and decay entering calculations of two-body current (MEC) effects: $C^{A}_{5}(Q^2)$ is the dominant $N\rightarrow\Delta$ transition form factor, and $\Lambda_{R}$ renormalizes the self energy of the $\Delta$ as described in Sec.~\ref{sec:FactorizationScheme}.}
    \label{fig:MECvariation}
\end{figure*}

\begin{table*}[t]
\centering
 \begin{ruledtabular}
 \begin{tabular}{cccc} 
   MiniBooNE & $0.2 < \cos\theta_{\mu} < 0.3$ & $0.5 < \cos\theta_{\mu} < 0.6$ & $0.8 < \cos\theta_{\mu} < 0.9$ \\\hline
   
SF Difference in $d\sigma_{\rm{peak}}$ $(\%)$  & $16.3$ & $17.1$ & $9.3$ \\

GFMC Difference in $d\sigma_{\rm{peak}}$ $(\%)$  & $18.6$ & $17.1$ & $12.2$ \\
 \end{tabular}
   \vspace{10pt}
 \begin{tabular}{cccc} 
 T2K & $0.0 < \cos\theta_{\mu} < 0.6$ & $0.80 < \cos\theta_{\mu} < 0.85$ & $0.94 < \cos\theta_{\mu} < 0.98$ \\\hline
 
SF difference in $d\sigma_{\rm{peak}}$ $(\%)$  & $15.3$ & $8.2$ & $3.3$ \\

GFMC difference in $d\sigma_{\rm{peak}}$ $(\%)$  & $15.8$ & $8.0$ & $4.6$ \\
 \end{tabular}
 \end{ruledtabular}
\caption{Percent increase in $\frac{d\sigma}{dT_{\mu}d\cos\theta_{\mu} }$ at the quasielastic peak between predictions using LQCD Bali et al./Park et al. $z$~expansion versus D2 Meyer et al. $z$ expansion nucleon axial form factor results. \label{table:quantFA}}
\end{table*}

\begin{figure*}[t]
    \centering
    \includegraphics[width=\textwidth]{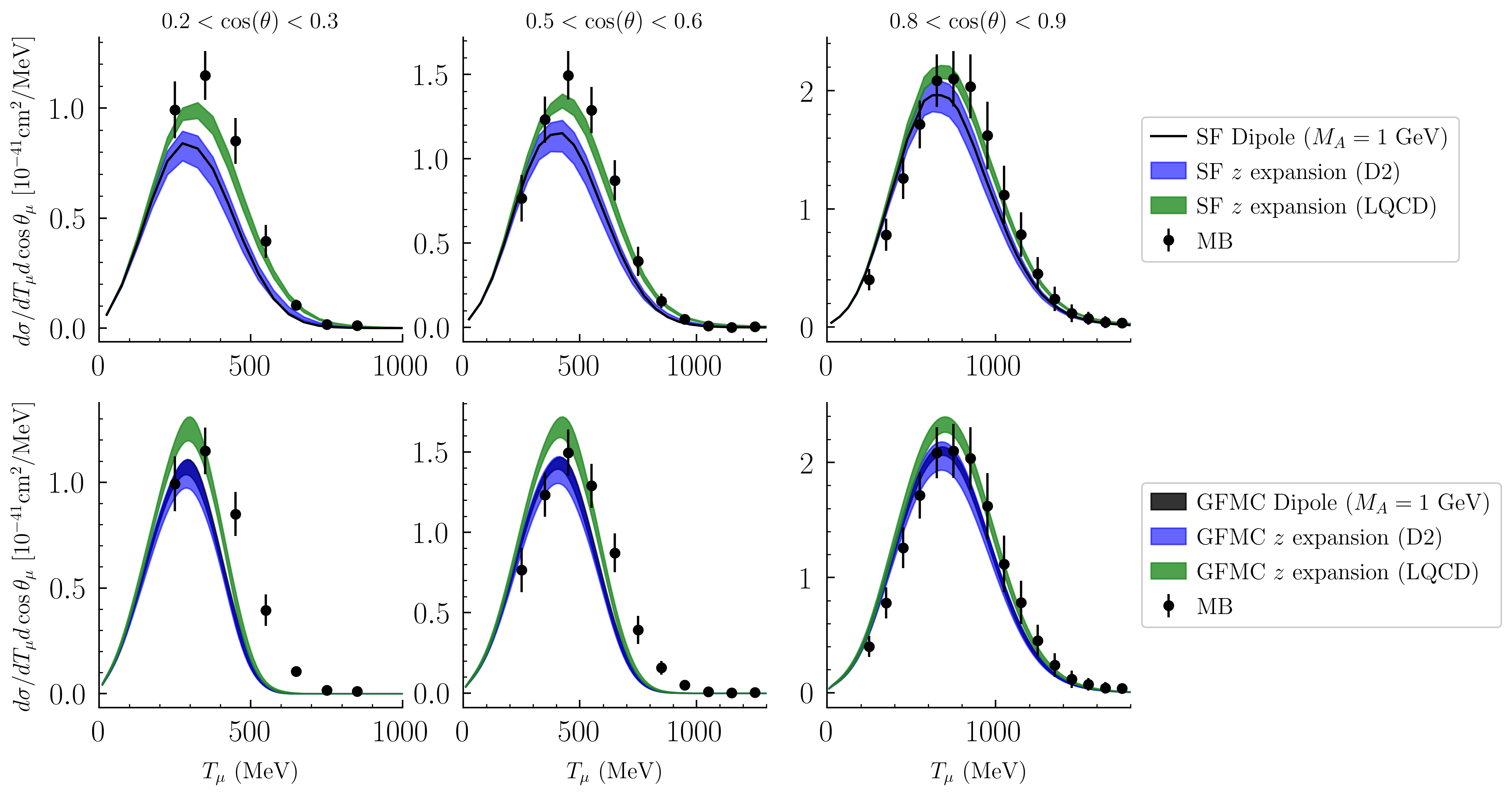}
    \caption{The $\nu_{\mu}$ flux-averaged differential cross sections for MiniBooNE. The top panel shows Spectral Function predictions in three bins of $\cos\theta_{\mu}$ with the D2 Meyer et al. $z$ expansion $F_{A}$ in blue, as well as the LQCD Bali et al./Park et al. $z$ expansion $F_{A}$ in green. The dipole parameterization with $M_A = 1.0$ GeV is shown without uncertainties as a black line. The lower panel shows GFMC predictions using the same set of axial form factors, although in the GFMC case systematic uncertainties including those arising from inversion of the Euclidean response functions are included in all results and the $M_A = 1.0$ GeV dipole form factor results are therefore shown as a black band.}
    \label{fig:datavstheory_FF_MB}
\end{figure*}

\begin{figure*}[t]
    \centering
    \includegraphics[width=\textwidth]{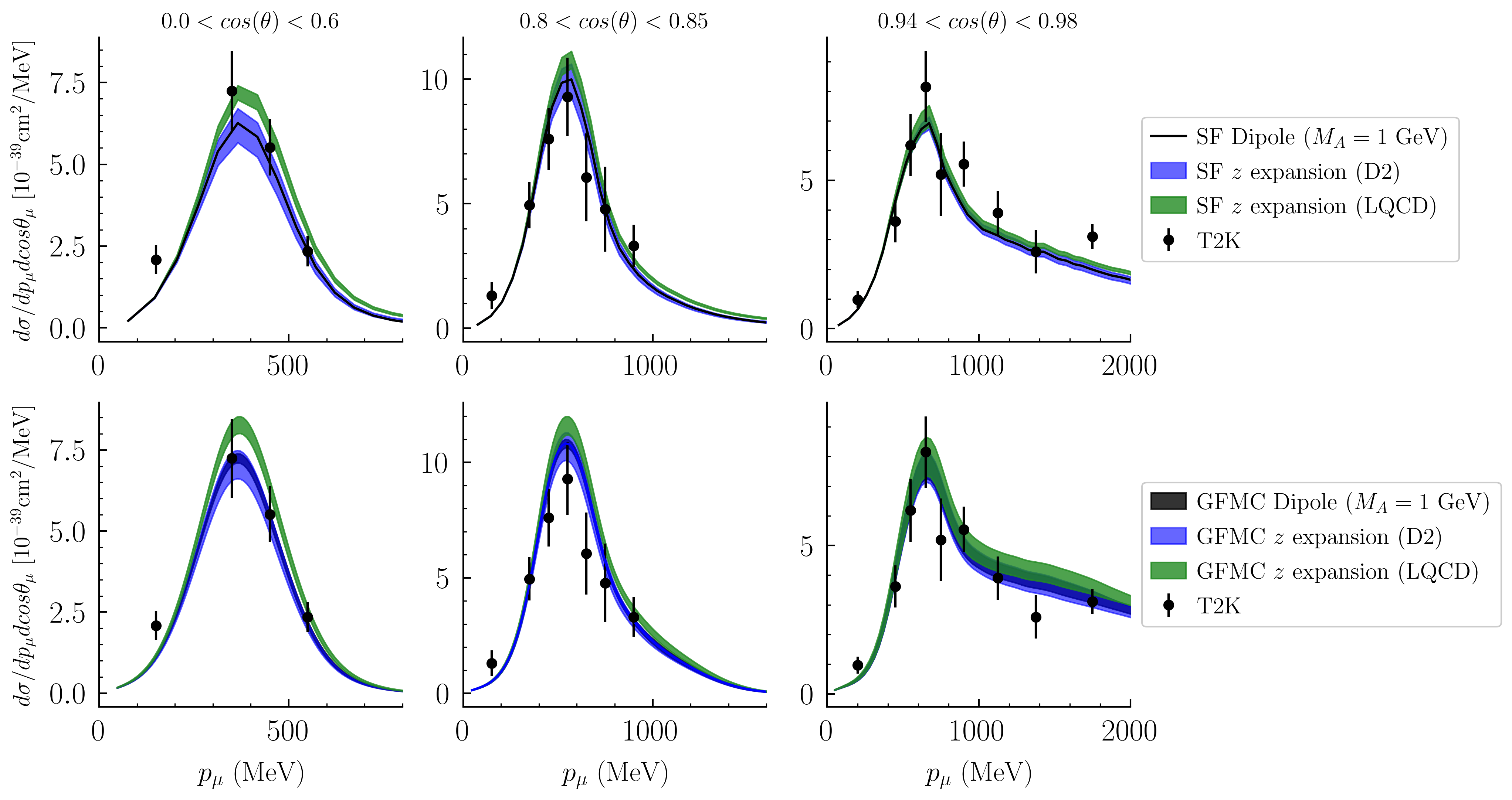}
    \caption{The $\nu_{\mu}$ flux-averaged differential cross sections for T2K. Details are as in Fig.~\ref{fig:datavstheory_FF_MB}.}
    \label{fig:datavstheory_FF_T2K}
\end{figure*}

\begin{figure*}
    \includegraphics[width=0.9\linewidth]{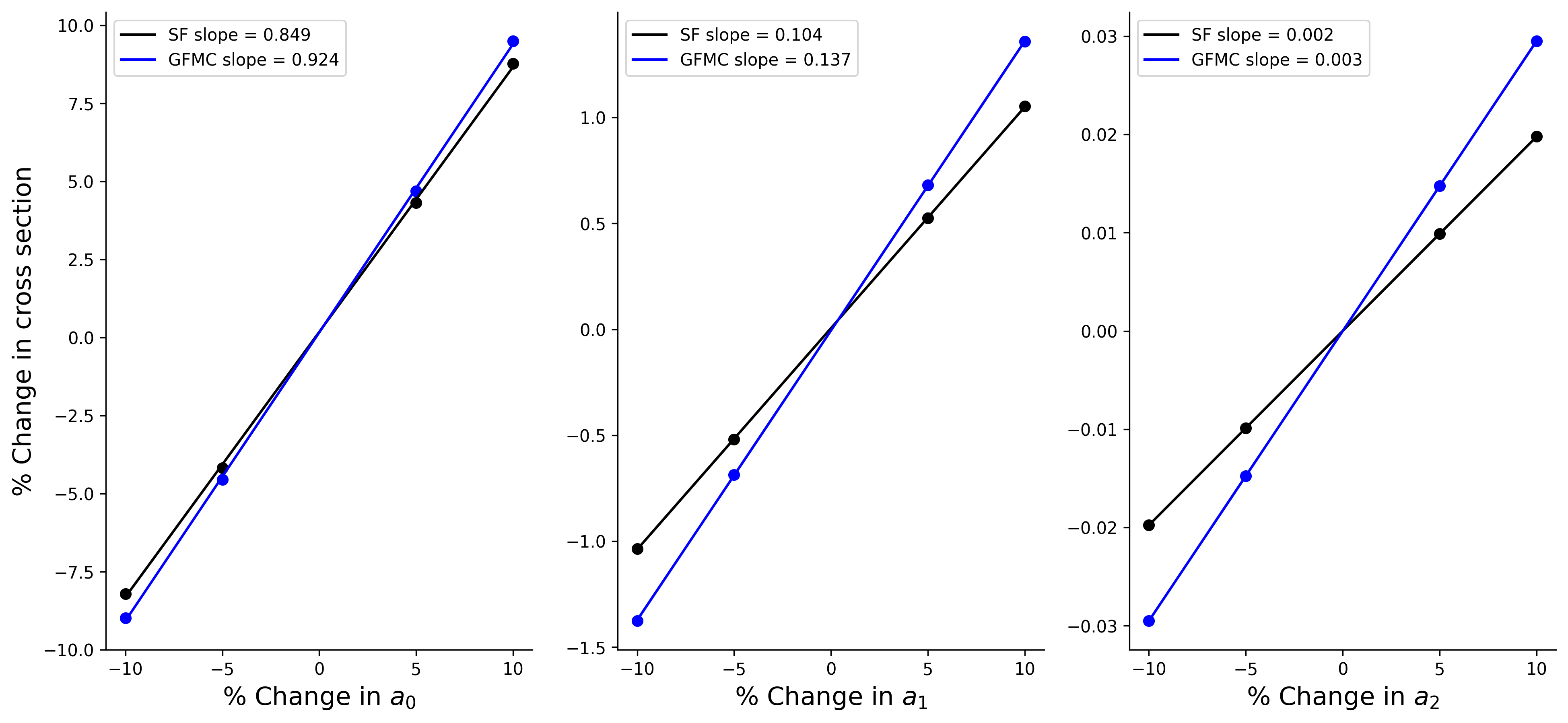}
    \caption{Percent change in peak value of MiniBooNE flux-averaged cross section for $0.5 < \cos\theta_{\mu} < 0.6$ vs. percent change in the z expansion parameters $a_{k}$. Results are shown for predictions using SF (black) and GFMC (blue) methods, including the slopes extracted from linear fits.}
    \label{fig:axial0506}
\end{figure*}

\begin{figure*}
    \includegraphics[width=\linewidth]{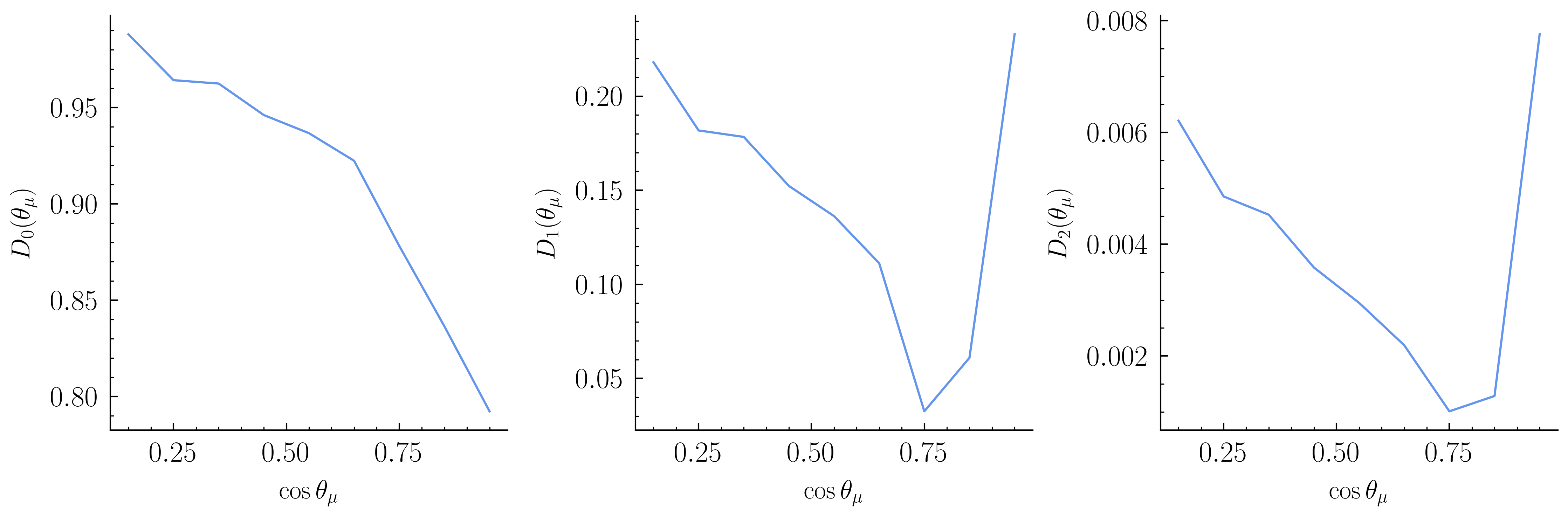}
    \caption{ The ratio between relative uncertainties in MiniBooNE cross sections and nucleon axial form factor $z$ expansion coefficients, $D_k(\theta_{\mu})$, vs muon scattering angle $\theta_\mu$ for $z$ expansion coefficients $a_k$ with $k\in \{0,1,2\}$. Results for $D_k(\theta_\mu)$ are calculated using the GFMC method with $T_\mu$ set equal to the peak value of the flux-averaged differential cross section for each $\theta_\mu$  and correspond to the slopes shown for particular $\cos\theta_\mu$ bins in Fig.~\ref{fig:axial0506}. 
    \label{fig:Dk_theta}  }
\end{figure*}

The two-body current contributions to flux-averaged cross sections are among the most unconstrained due to a lack of knowledge of the $N\rightarrow \Delta$ transition form factors and other nucleon resonance properties and two-nucleon correlations that enter the theoretical description of these contributions discussed in Sec.~\ref{sec:methods}. Empirical MEC models~\cite{Katori:2013eoa} and phenomenological models such as the Valencia model~\cite{Gran:2013kda} are used to describe two-body current effects in the neutrino event generators used for oscillation analyses at NOvA, MINERvA, and T2K, and neutrino experiments must resort to further tuning that significantly changes the shape and normalization of this 2p2h component in order to achieve satisfactory agreement between theory and data for flux-averaged cross sections~\cite{NOvA:2020rbg,MINERvA:2021wjs}. The degree to which any realistic model can reproduce the effect of these experimental tunes remains to be seen. To investigate the model dependence of two-body current effects, we have chosen two parameters from the SF two-body current calculation and varied them to investigate their effect on flux-averaged cross sections. These two parameters are $C^{A}_{5}(Q^2 = 0)$, which is the dominant coupling\footnote{Note that although $C_6(Q^2) > C_5(Q^2)$ for $Q^2 \lesssim 1$ GeV$^2$~\cite{Alexandrou:2010uk}, contributions to neutrino-nucleus cross sections from $C_6$ are suppressed by lepton masses and therefore sub-dominant. A relation between $C_6$ and $C_5$ analogous to Eq.~\eqref{eq:PCAC} is also predicted by leading order chiral perturbation theory. See Refs.~\cite{LlewellynSmith:1971uhs,Hernandez:2007qq} for more details.}  appearing in the $N\rightarrow\Delta$ vertex according to phenomenological models~\cite{Adler:1968tw,LlewellynSmith:1971uhs,Paschos:2003qr,Leitner:2008ue,Graczyk:2009qm,Hernandez:2007qq,Hernandez:2010bx} and pioneering LQCD calculations~\cite{Alexandrou:2006mc,Alexandrou:2010uk}, and $\Lambda_{R}$, which is a parameter that renormalizes the self energy of the $\Delta$. These parameters have been chosen because they affect the $\Delta$ piece of the two-body current, which we have seen provides the largest contribution, as well as because they are highly unconstrained. 

Each parameter was varied by $\pm 5,10\%$ and the effect on the flux-averaged cross section at the peak of the two-body contribution was computed. The effect can be seen in Fig.~\ref{fig:MECvariation} where we have plotted the percent change in the MiniBooNE cross section versus the percent change in each parameter for $0.5 < \cos\theta_{\mu} < 0.6$, $T_{\mu} = 325$ MeV. This was fit to a line so that as in Sec.~\ref{sub_sec:fa_param} the extracted slope is an estimate of the derivative of the cross section with respect to each parameter.
The derivative with respect to $C^A_5(0)$ is estimated to be 0.31, meaning that achieving a given cross-section uncertainty requires $C^A_5(0)$ to be known with $\lesssim 3$ times that uncertainty.
A similar though slightly smaller slope of 0.29 is found for $\Lambda_R$.
Current extractions of $C_5(0)$ rely on single pion production data from deuterium bubble chamber experiments~\cite{Radecky:1981fn,Kitagaki:1986ct,Kitagaki:1990vs}, and due to limited statistics model assumptions on the relations between $N \to \Delta$ transition form factors are typically included to reduce the number of fit parameters.
Depending on the model assumptions used, the resulting uncertainty on $C_5(0)$ is estimated to be 10-15$\%$ in the analysis of Ref.~\cite{Hernandez:2010bx}, with similar though slightly less conservative  uncertainties estimated in Refs.~\cite{Hernandez:2007qq,Graczyk:2009qm}.
Note that all of these analysis assume a dipole parameterization of $F_A$ as well as modified dipole parameterizations of $C_5^A$, and therefore it is possible that these uncertainties are still underestimated.
Even less is known about the uncertainty in determining $\Lambda_{R}$~\cite{Dekker:1994yc}. A $15\%$ variation in either $C_5^A(0)$ or $\Lambda_R$ changes the flux-averaged cross section by roughly $5\%$, and it will therefore be important to obtain more information on these parameters in order to achieve few-percent precision on cross-section predictions.

Focusing now on $F_{A}$, Figs.~\ref{fig:datavstheory_FF_MB} and~\ref{fig:datavstheory_FF_T2K} compare flux-averaged cross sections with different axial form factor determinations: a dipole form factor with $M_{A} = 1.0$ GeV, the D2 Meyer et al. $z$ expansion, and the LQCD Bali et al./Park et al. $z$ expansion. One can see that the LQCD $z$ expansion increases the normalization of the cross section across the whole phase space, with significantly more enhancement in the bins of low $\cos\theta_{\mu}$ corresponding to backward angles and higher $Q^2$. This is quantified in Table~\ref{table:quantFA}, which shows the percentage difference in the peak values of $\frac{d\sigma}{dT_{\mu}d\cos\theta_{\mu} }$ for the LQCD and D2 $z$ expansion results. The LQCD prediction increases the peak cross section between $10$-$20\%$, with the discrepancy growing at backwards angles.

To investigate the sensitivity of the flux-averaged differential cross section to variations in the axial form factor, derivatives of the MiniBooNE cross section with respect to the model-independent $z$ expansion parameters $a_k$ are computed as described in Sec.~\ref{sub_sec:fa_param}. Figure~\ref{fig:axial0506} shows the percent differences in flux-averaged cross sections evaluated at the quasielastic peak that have been computed using both GFMC and SF methods after independently varying each $a_k$ by $\pm 5,10\%$. The slopes of the resulting linear fits provide model-independent determinations of the sensitivity of the peak cross section to variations in $F_A$. It is clear that the impact of varying each $a_k$ decreases as $k$ increases, as expected since the contribution of each $a_k$ is suppressed by the $k$-th power of $z(Q^2) < 1$. In particular, a $10\%$ change in $a_0$ results in a $10\%$ change to the peak cross section, while a $10\%$ change in $a_1$ results in a $1\%$ change in the peak cross section, and $10\%$ variation of $a_k$ with $k \geq 2$ leads to sub-percent changes in the peak cross section. It is noteworthy that the uncertainty relations determined using GFMC and SF methods are reasonably consistent, with slightly larger slopes arising in GFMC calculations where the axial form factor appears in one- and two-body interference terms in addition to one-body contributions.
We have further studied the $\cos\theta_{\mu}$ dependence of these uncertainty relations using the GFMC method as shown in Fig.~\ref{fig:Dk_theta}. It can be seen that the uncertainty relation factor $D_k(\theta_\mu)$ defined in Eq.~\eqref{eq:chain_rule_coefficient}, which is equivalent to the slopes shown for particular $\cos\theta_\mu$ bins in Fig.~\ref{fig:axial0506}, decreases as $\cos\theta_{\mu}$ increases, except for $\cos\theta_{\mu} \geq 0.75$ for $a_1$ and $a_2$. These inflection points are merely due to the absolute values appearing in the definition of $D_k(\theta_{\mu})$. Achieving $1\%$ precision in $a_0$ and $10\%$ precision in $a_1$ is seen to be sufficient to ensure that axial form factor uncertainties lead to only $1\%$ uncertainties in MiniBooNE flux-averaged differential cross sections across the entire phase space. 
The D2 Meyer et al. results lead to $6\%$ precision on $a_1$ but only $10\%$ precision on $a_0$,\footnote{Note that $a_0$ is highly correlated with the other $a_k$ in this determination because $a_0$ is fixed by treating Eq.~\eqref{eq:sum0} as a constraint.} and therefore higher-precision determinations of the axial form factor are necessary for achieving few-percent cross-section uncertainties.
Current LQCD results quote higher precision on $a_0$~\cite{Park:2021ypf,Djukanovic:2022wru}, though the LQCD studies use different values of $t_0$ that prevent an immediate quantitative comparison with these uncertainty targets.
It would simplify comparisons and uncertainty quantification for future LQCD calculations to use the Meyer et al. value of $t_0 = -0.28$ GeV$^2$, which is chosen to optimize the rate of convergence of the $z$ expansion over the interval $0 \leq Q^2 \leq 1$ GeV$^2$~\cite{Meyer:2016oeg}.

\section{Conclusions}\label{sec:conclusion}

Precise theoretical calculations of neutrino scattering cross sections on target nuclei supplemented by reliable estimates of the associated uncertainties are essential for the success of the accelerator neutrino experimental program. A promising strategy for achieving this goal is to combine state-of-the-art nuclear many-body methods with LQCD calculations of few-nucleon observables, including single-nucleon form factors. In addition, it is critical to quantify the precision level required on such input parameters to compute neutrino-nucleus cross section with a given accuracy.

In this work, we analyze the impact of varying the nucleon axial form factors on flux-averaged charged-current neutrino-nucleus cross-sections as measured by the MiniBooNE and T2K experiments. Specifically, we consider a dipole form factor with $M_{A} = 1.0$ GeV, the D2 Meyer et al. $z$ expansion results fit to experimental deuterium bubble-chamber neutrinos scattering data~\cite{Meyer:2016oeg}, and the LQCD Bali et al./Park et al. $z$ expansion results~\cite{RQCD:2019jai,Park:2021ypf}, which are consistent with other recent LQCD calculations~\cite{Djukanovic:2022wru}. The nuclear many-body problem is solved using the GFMC and SF approaches, which rely on similar descriptions of the initial target state. For instance, the one- and two-nucleon SF of $^{12}$C are obtained from state-of-the-art VMC calculations that use as input the same nuclear Hamiltonian as the GFMC. 

For the first time, we validate the sum of one- and two-body current contributions obtained within the SF approach against flux-averaged electroweak cross sections. Using the D2 Meyer et al. form factors, we observe an overall good agreement with the MiniBooNE and T2K data --- there are tensions with the MiniBooNE data for small values of $\cos\theta_\mu$. At small $T_{\mu}$, these discrepancies might be due to the model dependent procedure adopted to subtract the pion-absorption contribution from the experimental data~\cite{MiniBooNE:2010bsu}.
Using instead the LQCD axial form factor in the SF calculations show an improved agreement with MiniBooNE data for these kinematics over the entire $T_{\mu}$ range. By contrast, the full SF predictions (with one- and two-body currents) using either choice of axial form factors appear to provide a good description of the T2K data for all the kinematics that we consider. Discrepancies between GFMC results and MiniBooNE data are seen for small values of $\cos\theta_\mu$, independent of the form factor of choice. They could be due to the nonrelativistic nature of the GFMC calculation in conjunction with the significant large-$E_\nu$ tail of the MiniBooNE flux. The GFMC results similarly show good agreement with T2K data using both form factor parameterizations. It is noteworthy that adopting the LQCD $z$ expansion results leads to a $10$-$20\%$ enhancement of the GFMC and SF cross section with respect to the D2 $z$ expansion across the whole phase space, with the largest effect seen in bins of low $\cos\theta_{\mu}$. We find that the SF and GFMC methods predict similar uncertainty-quantification results on the form factor dependence. The sensitivity analysis that we carry out in this work enables to determine an uncertainty targets for future LQCD axial form factor studies. \\

We observe that the GFMC and SF cross sections including one- and two-body contributions are consistent for the different kinematics and data sets considered. The differences in the strength and peak positions arise from different factors such as the nonrelativistic nature of the GFMC calculations and the static treatment of the $\Delta$ propagator in the nonrelativistic reduction of the two-body currents. The use of a factorization scheme that neglects final state interactions between the emitted particles and the spectator nucleons and the lack of one- and two-body current interference in the SF approach also contribute to the observed discrepancies. The inclusion of relativistic corrections in the kinematics of the GFMC electroweak responses is currently ongoing and it will be the subject of a future work; modeling one- and two-body interference effects yielding one nucleon emission in the final state requires some conceptual developments in the SF approach that will be investigated in the future.  

Reaction channels yielding the production of one or more pions in the final state have not been considered here. While including explicit pion degrees of freedom in the GFMC involves nontrivial difficulties~\cite{Madeira:2018ykd}, both resonant and non-resonant pion production mechanisms have been included in the SF method and validated against electron scattering data~\cite{Rocco:2019gfb}. We plan to carry out an analysis of the cross-section dependence on the input parameters needed to describe non-resonant $N\to N\pi$ transition amplitudes in a future work. 
Pioneering LQCD calculations of $N\to N\pi$ transition amplitudes have begun~\cite{Barca:2021iak} and will eventually supply robust QCD predictions to constrain these input parameters.
It is also noteworthy that there has been significant efforts to calculate nucleon properties using quark models involving approximate solutions to QCD Schwinger-Dyson and Fadeev equations~\cite{Barabanov:2020jvn,Proceedings:2020fyd} that can reproduce LQCD results for nucleon axial and pseudoscalar form factors~\cite{Chen:2021guo,ChenChen:2022qpy} and could provide valuable predictions of resonant and non-resonant pion production amplitudes that are not yet known precisely from LQCD.

In order to provide predictions for $^{16}$O and $^{40}$Ar needed for next-generation accelerator neutrino experiments, we will employ QMC methods that are capable to treat larger nuclear systems, e.g., the auxiliary field Diffusion Monte Carlo method~\cite{Schmidt:1999lik} and assess the uncertainty of our predictions. We plan to validate future predictions for $^{40}$Ar against accurate data from the Short-Baseline Neutrino Program as they become available and continue studying the nuclear many-body and nucleon form factor precision required to achieve the design cross-section uncertainty goals of DUNE.

\acknowledgments{
  We thank Gunnar Bali, Rajan Gupta, Andreas Kronfeld, and Aaron Meyer for helpful discussions.
This manuscript has been authored by Fermi Research Alliance, LLC under Contract No. DE-AC02-07CH11359 with the U.S. Department of Energy, Office of Science, Office of High Energy Physics  and Fermilab LDRD awards (N.S). The present research is also supported by the U.S. Department of Energy, Office of Science, Office of Nuclear Physics, under contracts DE-AC02-06CH11357 (A.L.), by the NUCLEI SciDAC program (A.L.), the DOE Early Career Research Program (A.L.), and by the Visiting Scholars Award Program of the Universities Research Association \#21-S-12. Quantum Monte Carlo calculations were performed on the parallel computers of the Laboratory Computing Resource Center, Argonne National Laboratory, the computers of the Argonne Leadership Computing Facility via the INCITE grant ``Ab-initio nuclear structure and nuclear reactions''.}

\bibliography{z_expansion_focus}

\end{document}